\documentclass[11pt,twocolappendix,apj]{emulateapj}
\usepackage{amssymb,amsmath}
\usepackage{graphicx}
\usepackage{natbib}
\usepackage{epstopdf}

\def\etal{{\it et al.\ }}

\def\p3m{P${}^3$M} 
\def\ap3m{AP${}^3$M} \def\-{{\em{---}}}

\def\tcc{t_{\rm cc}}
\def\cps{c_{\rm ps}}

\def\gsim{\;\rlap{\lower 2.5pt
\hbox{$\sim$}}\raise 1.5pt\hbox{$>$}\;}
\def\lsim{\;\rlap{\lower 2.5pt
\hbox{$\sim$}}\raise 1.5pt\hbox{$<$}\;}
\def\etal{{\it et al\ }}

\newcommand{\be}{\begin{equation}}  \newcommand{\ba}{\begin{eqnarray}}
\newcommand{\ee}{\end{equation}}  \newcommand{\ea}{\end{eqnarray}}

 \newcommand{\bi}{\begin{itemize}}
\newcommand{\ei}{\end{itemize}}

\def\lesssim{\mathrel{\hbox{\rlap{\hbox{\lower4pt\hbox{$\sim$}}}\hbox{$<$}}}}
\def\gtrsim{\mathrel{\hbox{\rlap{\hbox{\lower4pt\hbox{$\sim$}}}\hbox{$>$}}}}

\begin{document}

\title{The Launching of Cold Clouds by Galaxy Outflows II: \\ The Role of Thermal Conduction}

\author{Marcus Br\"uggen}
\affil{Universit\"at Hamburg, Hamburger Sternwarte, Gojenbergsweg 112, 21029, Hamburg, Germany}
\author{Evan Scannapieco}
\affil{School of Earth and Space Exploration,  Arizona State University, P.O. Box 871404, Tempe, AZ, 85287-1404}

\begin{abstract}

We explore the impact of electron thermal conduction on the evolution of radiatively-cooled cold clouds embedded in flows of hot and fast material, as occur in outflowing galaxies. Performing a parameter study of three-dimensional  adaptive mesh refinement hydrodynamical simulations, we show that electron thermal conduction causes cold clouds to evaporate, but it can also extend their lifetimes by  compressing them into dense filaments.  We distinguish between low column-density clouds, which are disrupted on very short times, and high-column density clouds with much-longer disruption times that are set by a balance between impinging thermal energy and evaporation. We provide fits to the cloud lifetimes and velocities that can be used in galaxy-scale simulations of outflows,  in which the evolution of individual clouds cannot be modeled with the required resolution. Moreover, we show that the clouds are only accelerated to a small fraction of the ambient velocity because compression by evaporation causes the clouds to present a small cross-section to the ambient flow.  This means that either magnetic fields must suppress thermal conduction, or that the cold clouds observed in galaxy outflows are not formed of cold material carried out from the galaxy.

\end{abstract}

\section{Introduction}

Galaxy outflows play a key role in  the history of galaxy formation \citep[e.g.][]{2011MNRAS.415...11D}, driving multiphase material from deep within the densest regions of galaxies out into the rarified fringes of the circumgalactic medium. Such powerful outbursts are well observed in high-surface density galaxies over a wide range of masses and redshifts \citep[e.g.][]{1999ApJ...513..156M, 2001ApJ...554..981P, 2002ASPC..254..292H, 2005ARA&A..43..769V}, including the ultraluminous infrared galaxies (ULIRGs) that host as much as half of the $z \geq 1$ star formation in the Universe \citep{2005ApJS..160..115R}.  Their impact is extensive: they are believed to set the correlation between galaxy stellar mass and interstellar medium metallicity,  \citep[e.g.][]{2004ApJ...613..898T, 2008ApJ...681.1183K,2010MNRAS.408.2115M},  they enrich the intergalactic medium with metals \citep[e.g.][]{2003ApJ...597L..97P, 2005ApJ...634L..37F, 2010ApJ...721..174M, 2014ApJ...786...54P, 2015MNRAS.450.2067T}, and they help determine the number density of faint galaxies \citep[e.g.][]{SFM02,Bens+03}.

At the same time, galaxy outflows are notoriously difficult to simulate \citep[e.g.][]{2011ApJ...731...11C, 2013MNRAS.430.3213B, 2015MNRAS.446.2125C}.  This is both because (i) the efficient cooling in the interstellar medium (ISM) makes it hard to model supernova feedback in galaxy-scale simulations, and (ii) the treatment of turbulent multiphase gas subject to radiative cooling, which can become thermally-unstable
\citep[e.g.][]{2013ApJ...763L..31S} and which puts high demands on the range of spatial and temporal scales involved in the problem.  Moreover, the observation of the multiphase ISM is difficult: the hot phases of the outflow have only been detected in a few instances \citep[e.g.][]{1990ApJS...74..833H, 2014ApJ...790..116V, 2014ApJ...781...55W}, and the cold phases of gas are often difficult to interpret.  Nevertheless the kinematics of neutral, atomic outflows has been studied in many cases through absorption line as well as emission line measurements of Ly$\alpha,$  H$\alpha$, H$\beta$ and well as doublet lines such as [N II]  6549, 6583, [O II] 3726, 3729, and [O III] 4959, 5007 \citep[e.g.][]{2004ApJ...610..201S, 2005ApJS..160..115R, 2009ApJ...692..187W, 2012ApJ...759...26E, 2013ApJ...774...50K,2015ApJ...809..163A}.   

In \cite{2015ApJ...805..158S}, hereafter called Paper I, we performed a series of three-dimensional hydrodynamic simulations aimed at developing a better understanding of the cold-cloud  hot-medium interactions needed to relate such observations to the evolution of galaxy outflows.  The simulations were carried out on adaptive grids using the FLASH code \citep{2000ApJS..131..273F}, and tracked the evolution of spherical, $T\approx 10^4$ K clouds impacted by hot winds of varying temperatures and speeds. For these simulations, we included optically-thin radiative cooling and designed the grid such that it tracked the clouds closely, enabling us to study their entire evolution, even for long disruption times. 

From these calculations, we found that the Mach cone that forms around the cloud both damps shear instabilities and produces a streamwise pressure gradient that stretches the cloud. As a result, cold clouds in highly supersonic outflows can live substantially longer than in subsonic conditions.  The calculations also followed a number of simple scaling laws for the cloud survival time and acceleration rate. From these, we could determine that clouds can travel out to distances of about 40 times the cloud radius but not much farther before becoming disrupted. Hence, it is difficult to associate $T\approx10^4$ K gas at distances of tens of kpc with $T\approx 10^4$ K gas leaving the driving region of the starburst-driven outflow unless an additional physical process acts to preserve the cloud for longer.

In this second paper in this series, we focus on the impact of one such key physical process: thermal conduction, which occurs as rapid electrons from the hot surrounding medium move into the  cold cloud.    Previous investigations on the effect of this process on multiphase gas dates back several decades. In seminal work on thermal instabilities, \cite{1965ApJ...142..531F} identified a characteristic length scale below which thermal instabilities are suppressed by conduction. This length scale stems from a basic distinction between radiative cooling, which increases proportional to volume, and conductive heat flux, which increases proportional to area.  The resulting length scale \citep[termed Field length by][]{1990ApJ...358..392M}  is given by $\lambda_{\rm F}\equiv\sqrt{(\kappa(T) T/[n^2 \Lambda(T)]}$, where $\kappa(T)$ is the rate of thermal conduction and $\Lambda(T)$  is the equilibrium cooling function.  In other early work, \cite{1973ApJ...179..469G} investigated the balance between cosmic-ray heating and cooling and found that thermal conduction plays a role in determining cloud sizes, and  \cite{1977ApJ...211..135C} computed the analytic mass loss rate of a nonradiating spherical cloud embedded in a hot plasma,  both for classical conduction, as well as saturated conduction, in which the electrons stream into the clouds at their thermal speed.

More recently, \cite{1990ApJ...358..392M} and \cite{1990ApJ...358..375B} worked out analytic solutions for steady evaporation and condensation of isolated cold clouds in hot plasmas, and found the conditions under which a steady state is reached.   In related work, \cite{1993ApJ...404..625D} provided analytic solutions for the saturated (flux-limited) thermal evaporation of spherical clouds allowing the thermal conductivity to change continuously from a diffusive to a saturated form, in a manner usually employed only in numerical calculations. \cite{2007MNRAS.382.1481N} revisited the analytic treatment of cold clouds in hot media and argued that thermal evaporation of cold gas by the hot coronae of galaxies can explain the surface brightness properties of elliptical galaxies and  their active galactic nuclei. \cite{2015arXiv150701951} analytically modeled the dynamics of cool clouds in hot winds including thermal conduction, varying the parameters of the problem, and obtaining general constraints on ram pressure acceleration  by comparing velocities, column densities, and temperatures with observations

From a numerical perspective, \cite{2007A&A...475..251V, 2007A&A...472..141V} simulated the conductive evaporation of spherical clouds embedded in a stationary and a moving ambient medium, respectively. \cite{2005MNRAS.362..626M} also presented high-resolution hydrodynamical models of warm ionized clouds embedded in a wind, and compared to the OVI and X-ray properties of galaxy outflows, but their simulations were limited to relatively short times and yielded fairly short lifetimes of the warm clouds.   \cite{2008ApJ...678..274O} performed 2.5-D MHD simulations of a shock impacting on an isolated gas cloud, including anisotropic thermal conduction and radiative cooling. Other efforts to model cloud-wind interactions in the interstellar medium include \cite{1994ApJ...420..213K}, \cite{1994ApJ...433..757M}, \cite{1999ApJ...527L.113G}, \cite{2002A&A...395L..13M, 2004ApJ...604...74F}, \cite{2005A&A...444..505O}, \cite{2008ApJ...680..336S}, \cite{2012ApJ...756..102P}, \cite{2013ApJ...766...45J}, \cite{2013ApJ...774..133L}, \cite{2015MNRAS.449....2M}, and  \cite{2015arXiv151005478P}.

Most related to our current work is  \cite{2005A&A...444..505O}. Using the same code as we do, the FLASH code, they investigated the competition between radiative cooling and thermal conduction during the evolution of spherical clouds exposed to a planar shock wave. In 3D simulations, they explored two cases, albeit with very limited spatial resolution.  In case (i) with $t_{\rm cool} < t_{\rm evap}$ and an exterior Mach number of 30,   radiative losses dominated the evolution of the shocked cloud, which fragmented into cold, dense, and compact filaments surrounded by a hot corona that was ablated by the thermal conduction.  In case (ii)  with $t_{\rm cool} > t_{\rm evap}$  and an exterior Mach number of 50, thermal conduction dominated the evolution of the shocked cloud, which evaporated in a few cloud crushing times. In both cases, they found that thermal conduction was  effective in suppressing the hydrodynamic instabilities that would otherwise develop at the cloud interface. 

Here we adopt a similar approach to investigate lower external Mach numbers, higher density ratios, and the dependence on cloud size with high-resolution calculations that continue for many dynamical times where necessary.   Building on our work in Paper I, we conduct a suite of 3D FLASH simulations that include radiative cooling and electron thermal conduction, and span the range of conditions observed in galaxy outflows.  These simulations allow us to both obtain scalings that can be used in galaxy-scale simulations of outflows as well as  draw general conclusions that can be used when interpreting observations.  Note that in this study we neglect both the dynamic effects associated with magnetic fields, as well as the suppression of  thermal conduction perpendicular to field lines, representing two important caveats for the conclusions described below.

The structure of this paper is as follows.  In \S2  we describe the physics of the interaction between cold clouds and the ambient hot wind with a focus on the expected role of thermal conduction.  In \S 3, we describe the details of our simulation including the setup and the adaption of the grid. In \S4, we present the timescales for cloud disruption along with a physical model for the observed scalings.   In \S5 we give further details as to the morphology and evolution of the interactions, and in  \S6 we turn our attention to the acceleration of the cold clouds, which we find to be much less efficient than in the case neglecting conduction.  Numerical convergence is examined in \S7, and in \S8 we summarize our results and their implications for the launching of cold clouds by galaxy outflows.  An appendix describes the properties of strong shocks in the presence of electron thermal conduction.

\section{The Physics of Cold Clouds}

\subsection{Cold Cloud Disruption and Cooling}

As discussed in greater detail in Paper I, there are three important timescales that determine the evolution of a cloud overtaken by a hot, exterior flow.  The most relevant of these timescales is the cloud crushing time, or the approximate timescale for the shock induced in the cloud to travel a distance equal to the cloud radius, $R_{\rm c}.$   This is given by $\tcc  = \chi_0^{1/2} R_{\rm c}/v_{\rm h}$,  where $v_{\rm h}$ is the velocity of the hot, exterior medium, and $\chi_0$ is the initial ratio of the cloud density to the density of the initial exterior medium, which for a cloud in pressure equilibrium is also equal to the ratio of the exterior temperature to the cloud temperature \cite[e.g.][]{1994ApJ...420..213K}.

A second timescale is the cooling time behind the shock,
$t_{\rm cool} \equiv [3/2 n_{\rm c} k_{\rm B}T_{\rm ps}/(\Lambda(T_{\rm ps}) n_{\rm e,c} n_{\rm i,c})],$ where $T_{\rm ps}$ is the post-shock temperature, $\Lambda(T)$ is again the radiative cooling function, $k_{\rm B}$ is the Boltzmann constant, and $n_{\rm c},$ $n_{\rm i,c},$ and $n_{\rm e,c},$  are the total, ion, and electron number densities within the cloud, respectively. The ratio of the cooling time to the cloud crushing time is
\be
\frac{t_{\rm cool}}{t_{\rm cc}} =  \frac{3 n_{\rm c} k_{\rm B} T_{\rm ps} v_{\rm h}}{ 2 \Lambda(T_{\rm ps}) n_{\rm e,c} \chi_0^{1/2}}  \frac{1}{n_{\rm i,c} R_{\rm c}} = \frac{N_{\rm cool}}{n_{\rm i,c} R_{\rm c}},
\ee
where $N_{\rm cool} \equiv \frac{3 k_{\rm B} T_{\rm ps} v_{\rm h} n_c}{ 2 \Lambda(T_{\rm ps}) \chi_0^{1/2} n_{\rm e,c}}$ is a column density that is purely a function of the velocity of the transmitted shock, at least to the extent that the velocity of the transmitted shock can be approximated as $v_{\rm h}/\chi_0^{1/2}$. 
 
A third timescale is  the characteristic time for disruption by the Kelvin-Helmholtz (KH) instability. In the linear, subsonic case, this is directly proportional to the cloud crushing timescale \citep{1961hhs..book.....C}. In the nonlinear, supersonic case however, the growth rate of the boundary layer is $\Delta  v \approx   0.2 \, v_{\rm h} \, [1 +4 (\gamma-1) M_{\rm c}^2]^{-1/2}$ with $M_{\rm c}$ the ratio of $v_{\rm h}$ with the  minimum of the sound speeds of the shearing fluids (Slessor \etal 2000), which in this case is the sound speed of the cloud, $c_{\rm s,c}$.  Thus the shear layer will grow to the size of the cloud in a timescale $t_{\rm KH} \approx 5 (R_{\rm c}/v_{\rm h})  [1+ 4 (\gamma-1) M_{\rm c}^2]^{1/2}$ which for high Mach number gives  $t_{\rm KH, nl} \approx 10 R_{\rm c}/c_{\rm s,c}$.  This means that $t_{\rm KH, nl}/t_{\rm cc}  \approx 10 v_{\rm h}/\chi_0^{1/2} c_{\rm s,c}  \approx 10  M$  where $M$ is the Mach number of the flow relative to the {\em exterior} sound speed.  This somewhat counterintuitive result is due to the fact that the cloud and the exterior medium are initially in pressure equilibrium, then the Mach number of the transmitted shock that propagates through the cloud is the same as the Mach number of the exterior shock.  This suggests that in cases in which the KH instability is the primary cause of cloud disruption, clouds will be preserved for longer if they are impacted by gas at higher exterior Mach numbers.  However, this instability is not the primary cause for disruption in the case with efficient electron thermal conduction.

\subsection{Thermal Conduction}

In the absence of magnetic fields, electron thermal conduction  leads to the following equations for the evolution of mass, momentum, and energy
\begin{eqnarray}
\partial_t \rho + \nabla\cdot\left(\rho {\bf u}\right)&=&0 ,\\
\rho \left[ \partial_t  {\bf u} +  ({\bf u} \cdot \nabla) {\bf u}\right]&=&- \nabla p,  \\
\partial_t E + \nabla\cdot\left[E {\bf u}\right]  & = & - 
 \nabla\cdot \left( p {\bf u} \right)  - n^2 \Lambda(T) +  \nabla {\bf q}  ,
\label{eqnmhd}
\end{eqnarray}
where $\rho$, ${\bf u}$, $p= k_B T \rho/(\mu m_{\rm p})$, and $E= p/(\gamma-1)  + \frac{1}{2} \rho\left|{\bf u}\right|^2 $ denote the density, velocity, pressure, and total (internal and kinetic) energy density,  $\mu$ is the average particle mass in units of the proton mass $m_{\rm p}$,  $\Lambda(T)$ is again the radiative cooling function, and 
\be
{\bf q} = {\rm min} \, \begin{cases}
\kappa(T) \nabla T\\
0.34 n_e k_{\rm B} T  c_{\rm s,e},
\end{cases} 
\label{eq:thermalcond}
\ee
 \citep{1977ApJ...211..135C}, where $\kappa(T) = 5.6 \times 10^{-7} T^{5/2}$ erg s$^{-1}$ K$^{-1}$ cm$^{-1}$  and  $c_{\rm s,e} = (k_{\rm B} T/m_{\rm e})^{1/2}$ is the isothermal sound speed of the electrons in the exterior gas, with $m_{\rm e}$  the mass of the electron. Throughout this study, we will assume that the electrons and ions have the same temperature. Furthermore, the mean free path of electrons into the cloud is $\lambda_{\rm i} =  1.3 \times 10^{18} {\rm cm}^{-2} T_7^2/n_{\rm i,c}$, where $n_{\rm i,c}$ is the ion density within the cloud and $T_7 = T/10^7$ K. Note that these equations are invariant under a transformation in which 
\be
{\bf x} \longrightarrow \alpha {\bf x}, \qquad
{\bf t} \longrightarrow \alpha {\bf t}, \,   \qquad {\rm and} \qquad
\rho \longrightarrow \rho/\alpha. 
\ee
In other words, if one works in units of the cloud crushing time, the evolution of the cloud will depend only on the product of its size and its density,  greatly reducing the parameter space that we need to simulate.

Our assumption of a sharp transition between the edge of the cloud and the hot medium means that we are interested in the  case in which $|T/\nabla T| < \lambda_{\rm i},$ such that conduction occurs in the saturated regime.
Furthermore, if the column depth of the cloud is less than $\lambda_{\rm i} n_{\rm i,c} =  1.3 \times 10^{18} {\rm cm}^{-2} T_7^2$, the conducted energy will be deposited over the full volume of the cloud,  
whereas if the column depth of the cloud  is greater than this value it will be deposited in a region with a volume $\approx 4 \pi R_{\rm c}^2 \lambda_{\rm i},$ such that the total heat flux from the hot medium is given by 
$4 \pi R_{\rm c}^2  \times 0.34 n_{\rm e} k_{\rm B} T  c_{\rm s,e}$.   For $T=10^7$ K we have an isothermal sound speed of $\approx$ 12,000 km s$^{-1}$  such that $0.34 c_{\rm s,e} k_B T$ = 0.5 $T_7$ erg cm/s.

In both these cases we can compare the rate of energy loss through radiation in the volume impacted by conduction to its heating rate.
For high column depth clouds with $n_{\rm i,c} R_{\rm c}>  1.3 \times 10^{18} {\rm cm}^{-2} T_7^2$,  the ratio of these rates is given by
 \be
\frac{ \dot e_{\rm heat}}{\dot e_{\rm cool}}= 
\frac{4 \pi R_{\rm c}^2  \,0.34 c_{\rm s,e} n_{\rm e} k_B \, T} 
{4 \pi R_{\rm c}^2   \Lambda n_{\rm e,c} n_{\rm i,c}    \lambda_{\rm i}}
\approx  \frac{0.44}{\Lambda_{-21}} \left ( \frac{\chi}{1000}\right )^{-1} T_7^{-1/2},
\ee
where initially when the cloud is in pressure equilibrium with the exterior medium,  $\chi = 1000 T_7$, and $\Lambda_{-21}=\Lambda/10^{-21}$erg cm$^3$.
This means that if the exterior medium is relatively hot, the strong density contrast between the pressure confined clouds and the 
exterior medium will always ensure that cooling is faster than conduction within the conducting region.   
For low column depths with $n_{\rm i,c} R_{\rm c} < 1.3 \times 10^{18} {\rm cm}^{-2} T_7^2$ on the other hand, 
\begin{align}
&\frac{ \dot e_{\rm heat}}{\dot e_{\rm cool}}= 
\frac{4 \pi R_{\rm c}^2  \,0.34 c_{\rm s,e} n_{\rm e} k_B \, T} 
{(4/3) \pi R_{\rm c}^3   \Lambda n_{\rm e,c} n_{\rm i,c} } \nonumber \\
&\approx  \frac{0.44}{\Lambda_{-21} } \left ( \frac{\chi}{1000}\right )^{-1} T_7^{3/2}
\left ( \frac{4 \times 10^{18} {\rm cm}^{-2} }{n_{\rm i,c} R_{\rm c}}\right ),
\label{eq:heatcool}
\end{align}
implying that the smaller the column, the higher the heating over cooling ratio.

\section{Methods}
As in Paper I, we have carried out a suite of simulations that span the range of conditions expected in galaxy outflows.    The simulations are conducted  using  FLASH version 4.2, a multidimensional hydrodynamics code \citep{2000ApJS..131..273F}  that solves the fluid equations on a Cartesian grid with a directionally-split  Piecewise-Parabolic Method (PPM) (Colella \& Woodward 1984). In this study, we use the unsplit flux solver with a predictor-corrector type formulation based on the method presented in \cite{2013JCoPh.243..269L}, which achieves second-order solution accuracy for smooth flows and first-order accuracy for both space and time. For the interpolation schemes, we enable an adaptively varying-order reconstruction scheme reducing its order to first-order depending on monotonicity constraints. Moreover, we set the Courant-Friedrichs-Lewy timestep condition to 0.2.
 
For our initial conditions, we set up a spherical cloud of radius $R_{\rm c}=100$ pc and an initial temperature of $10^4$ K and a fiducial mean density of $\rho_{\rm c} = 10^{-24}$ g cm$^{-3}$, such that $R_{\rm c} n_{\rm i,c} = 1.5 \times 10^{20}$ cm$^{-2}.$  In all the simulations presented here, the computational grid covered a physical box of $-8 R_{\rm c} \times 8 R_{\rm c}$ parsecs in the $x$ and $y$ and directions and $-4R_{\rm c}$ to $8 R_{\rm c}$ parsecs in the $z$ direction, i.e. the direction of the ambient flow.  Exterior to the cloud, the initial velocity and sound speed of the gas were assumed to be $v_{\rm h}$, and $c_{\rm h}$ as given below. Finally, the density was determined by pressure equilibrium with the cold cloud. Note that for $T=10^4$ K and $\rho_{\rm c} = 10^{-24}$ g cm$^{-3}$ the Jeans length is $c_{\rm c}(G\rho)^{-1/2}\approx$ 2 kpc, implying that the blobs are pressure confined rather than gravitationally bound, and hence we do not include gravity in our simulations. We also monitor the Jeans length and find that it never falls below $\approx$ 300 pc.
 
In Table~\ref{tab:disruption}, we have listed the runs that are discussed in this paper.  They span 8 of the 11 cases studied in Paper I, which were chosen to cover the range of conditions encountered in starburst-driven galaxy outflows.  Here the lowest Mach number runs approximate conditions found at or near the base of the outflowing region, whereas the highest Mach number cases approximate conditions found at large distances from the galaxy, where adiabatic cooling has greatly decreased the temperature of the hot wind material.  In this study, we have chosen three runs with a Mach number of $\approx 1$ and three different ambient temperatures, three runs with a Mach number of $\approx 3.5$ and  three different ambient temperatures, a single Mach $\approx 6.5,$ run, and a single Mach $\approx 11.5$ run that was also repeated for a case in which the cloud and exterior density were both increased by a factor of 10. Furthermore, the Mach 3.8 run with $\chi_0=1000$, $v_{\rm h}=1700$ km/s was repeated at higher and lower resolution,  and the Mach 3.6 run with $\chi_0=3000$, $v_{\rm h}=3000$ km/s was repeated for cloud sizes of 10 and 1 pc, yielding altogether 13 production runs. Below we refer to the runs by naming them as $\chi \#v\#$.

At the lower $-z$ boundary of the simulation domain, material was continuously added to the numerical grid with the initial values of velocity, temperature, and density. On the $x$ and $y$ boundaries, as well as at the upper $+z$ boundary we employed the FLASH zero-gradient ``diode" boundary condition, which does not allow gas to flow back into the computational volume.  The large $x$ and $y$ sizes of the simulation domain ensured that this zero-gradient boundary condition does not influence the shape of the shock, leading to artifacts if the simulation domain is too small.   Moreover, a sufficient distance has to be left in front of the cloud, such that the bow shock remains within the computational domain, and a sufficient distance has to be left in the wake of the cloud in order to capture the filamentary evolution described below.
 
As in Paper I, we tracked the cloud using a massless scalar and computed its center-of-mass (COM) position ${\bf x_{\rm c}}$, velocity  ${\bf v_{\rm c}}$, and extent in the $x$, $y$, and $z$ directions, via the mass-weighted average values of ${\rm abs}(x-x_{\rm c}),$ ${\rm abs}(y-y_{\rm c}),$ and ${\rm abs}(z-z_{\rm c})$. As explained in Paper I, once per $t_{\rm cc},$ we moved the grid from a frame in which the cloud was drifting out towards the $+z$ boundary, to a new frame in which the cloud was drifting (much more slowly) towards the $-z$ boundary. In this way, we were able to keep the cloud COM close to $z=0$ at all times.

The simulation box was divided into $8\times 8\times 6$ blocks of $8^3$ cells each and we allowed for 5 levels of refinement yielding a minimum cell size of $\Delta x  = R_{\rm c}/64 = 4.8 \times 10^{18} {\rm cm}.$ We automatically refined on density and temperature discontinuities, and in order to obtain the most reliable results possible, we also maintained high-resolution throughout the simulation in the regions within the cloud and immediately around it.  Thus, when $t \leq 2 t_{\rm cc},$ we increased the refinement to the maximum within a cylindrical region with ${\rm abs}( z) \leq 1.5 R_{\rm cloud}$ and a distance from the $z$ axis $\leq 1.5 R_{\rm cloud}$ (see Paper I).  Conversely, to avoid unnecessary refinements in regions of the simulation not important to the cloud evolution, we automatically enforced de-refinement if the distance to the $z$ axis was greater than $> 3\, R_{\rm c},$ or $3 \times$  the instantaneous $x$ extent of the cloud {\em and} both ${\rm abs}(z)$ and ${\rm abs}(z-z_{\rm c}) > 3\, R_{\rm c}$.

Optically thin cooling was implemented using the tables compiled by \cite{2009MNRAS.393...99W} from the code CLOUDY \citep{1998PASP..110..761F} assuming solar metallicity.   Within the cooling routine, sub-cycling was implemented \citep{2010ApJ...718..417G} and we introduced a cooling floor at $T=10^4$ K.

The diffusion equation is an implicit equation and we solved it using the general implicit diffusion solver in FLASH. The implicit diffusion module uses the HYPRE\footnote{https://computation.llnl.gov/casc/hypre/software.html} linear algebra library to solve the discretized diffusion equation. Saturated thermal conduction was implemented by using a flux limiter that modifies the diffusion coefficient to vary smoothly up to some maximum flux, where we have used the Larsen flux limiter \citep{2000JQSRT..65..769M}.

The runs were carried out on the Stampede cluster at the Texas Advanced Computing Center (TACC) and on the JUROPA/JURECA supercomputers at the Forschungszentrum J\"ulich, Germany. Both machines are made up of dual 8-core Intel Xeon E5 processors. Typical production runs were carried out on 960 cores and took $\approx$ 20,000-60,000 CPU hours each. 

\section{Disruption Timescales}

\begin{figure*}[t!]
\begin{center}
\includegraphics[width=1.0\textwidth]{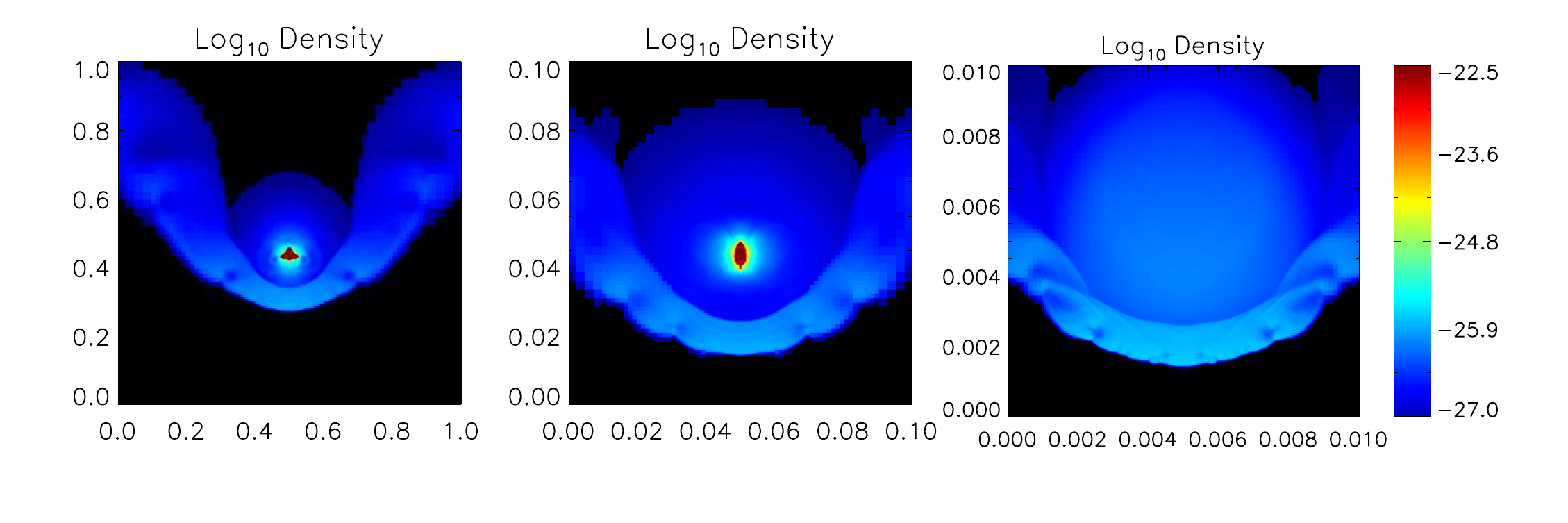}
\vspace{-0.4in}
\caption{Density slices through the centre of the computational box at 1 $\tcc$ with $\chi_0=3000$, $T=10^7$ K and $v_{\rm h}=3000$ km/s for three different cloud sizes, i.e. three different column densities of the cloud. From left to right we show a cloud size of 100 pc, 10 pc and 1 pc.}
\label{fig:column}
\end{center}
\end{figure*}

Fig.~\ref{fig:column} illustrates what occurs in the case in which the initial column depth, $n_{\rm i,c} R_{\rm c},$ approaches and drops below $1.3 \times 10^{18} {\rm cm}^{-2} T_7^2.$ In this figure we show density slices at 1 $\tcc$ from runs with $\chi_0=3000$, $T_{\rm h}=10^7$ K, $v_{\rm h}=3000$ km/s and three different cloud sizes. 
From left to right the slices correspond to cloud sizes of 100 pc, 10 pc and 1 pc, corresponding to $n_{\rm i,c} R_{\rm c}$ values of $1.5 \times 10^{20}$ cm$^{-2}$, $1.5 \times 10^{19}$ cm$^{-2}$ and $1.5 \times 10^{18}$ cm$^{-2}$, respectively.  In the larger column runs, the cooling rate exceeds the heating rate
in the outer layer of the cloud, leading to long phase of controlled evaporation as described below.  In the smallest cloud case, not only is $\dot e_{\rm heat} \approx \dot e_{\rm cool}$ but conduction is into the entire cloud decreasing $\chi$ from its initial value.   This causes rapid heating and the cloud is evaporated completely, well before a single cloud crushing time has passed.

As this rapid disruption by runaway heating will occur in any case in which the cloud column is less than a mean free path, in the remainder of this study we focus on the higher-column density case, fixing $R_{\rm c} = 100$ pc. In Tab.~\ref{tab:disruption} we list the main parameters of these runs (Mach number, initial density ratio, velocity, ambient temperature, and ambient sound speed) as well as the times when the clouds reach 90\%, 75\%, 50\% and 25\% of their original mass (denoted as $t_{\rm 90}$, $t_{\rm 75}$, $t_{\rm 50}$ and $t_{\rm 25}$) and the velocities of the COM of the clouds at those times (denoted as $v_{\rm 90}$, $v_{\rm 75}$, $v_{\rm 50}$ and $v_{\rm 25}$).

\begin{table*}[htdp]
\footnotesize
\caption{Parameters and results of our simulations. Columns show the Mach number, initial density ratio, velocity, ambient temperature (in units of $10^6$ K), ambient sound speed, and column density (in units of cm$^{-2}$), as well as the times when the clouds only had 90\%, 75\%, 50\% and 25\% of their original mass (denoted as $t_{\rm 90}$, $t_{\rm 75}$, $t_{\rm 50}$ and $t_{\rm 25}$), and the velocities of the COM of the clouds at those times (denoted as $v_{\rm 90}$, $v_{\rm 75}$, $v_{\rm 50}$ and $v_{\rm 25}).$ 
All times are in units of cloud crushing times and all velocities are in units of km/s.}
\begin{center}
\begin{tabular}{|c|c|c|c|c|c|c|c|c|c|c|c|c|c|}
\hline
$M$ & $\chi_0$ & $v_{\rm h}$ & $T_{\rm h}$ & $c_{\rm h}$ & 
$n_{\rm i,c} R_{\rm c}$ &
$t_{\rm 90}$ & $t_{\rm 75}$ & $t_{\rm 50}  $ & $t_{\rm 25}$ & $v_{\rm 90}$ & $v_{\rm 75}$ & $v_{\rm 50}$ & $v_{\rm 25}$\\
%    & & (km/s) & ($10^6$K) & (km/s) & (cm$^{-2}$) & ($t_{\rm cc}$) & ($t_{\rm cc}$)& ($t_{\rm cc}$) &($t_{\rm cc}$)& (km/s) & (km/s) & (km/s) &(km/s) \\
\hline
1.00    &    1000  &    480  & 10   &   480  &  $1.5 \times 10^{20}$ & 0.46  &  1.71  &    3.41  &   6.12   &   3.60   &   7.04  &  8.00  & 10.8\\
1.03    &    3000  &    860  & 30   &   832  &  $1.5 \times 10^{20}$ &  0.12  &  0.64  &    1.14  &   1.91   &   7.35   &   8.89  &  12.7  & 19.6\\
0.99    &  10000  &  1500  & 100 & 1519  &  $1.5 \times 10^{20}$ &  0.09  &  0.42  &    0.75  &   1.27   &   8.62   &   13.7  &  12.2  & 15.8\\
3.80    &      300  &  1000  & 3     &   263  &  $1.5 \times 10^{20}$ &  2.60  &  7.57  &  16.0  & 26.0       &  26.3    &   37.1  &  61.4  & 110\\
3.54    &    1000  &  1700  & 10   &   480  &  $1.5 \times 10^{20}$ &  0.44  &  2.00  &    3.38  &   5.60   &   14.5   &   21.5  &  28.3  & 37.2\\
3.61    &    3000  &  3000  & 30   &   832  &  $1.5 \times 10^{20}$ &  0.20  &  0.80  &    1.34  &   2.03   &   20.4   &   35.1  &  44.0  & 40.9\\
6.46    &      300  &  1700  & 3     &   263  &  $1.5 \times 10^{20}$ &  0.69  &  3.94  &    9.19  & 15.2   &   26.4   &   47.6  &  65.6  & 79.7\\
11.4    &      300  &  3000  & 3     &   263  &  $1.5 \times 10^{20}$ &  0.27  &  0.83  &    2.88  &    6.83  &   32.2   &   83.3  &  132   & 176\\
11.4    &      300  &  3000  & 3     &   263  &  $1.5 \times 10^{21}$ &  0.91  &  3.66  &  15.5  &  28.3      &   54.9   &   95.0  &  152   &  227\\            
\hline  
\end{tabular}
\end{center}
\label{tab:disruption}
\end{table*}%

\begin{figure*}
\begin{center}
\includegraphics[trim=0 0 0 0, width=\textwidth]{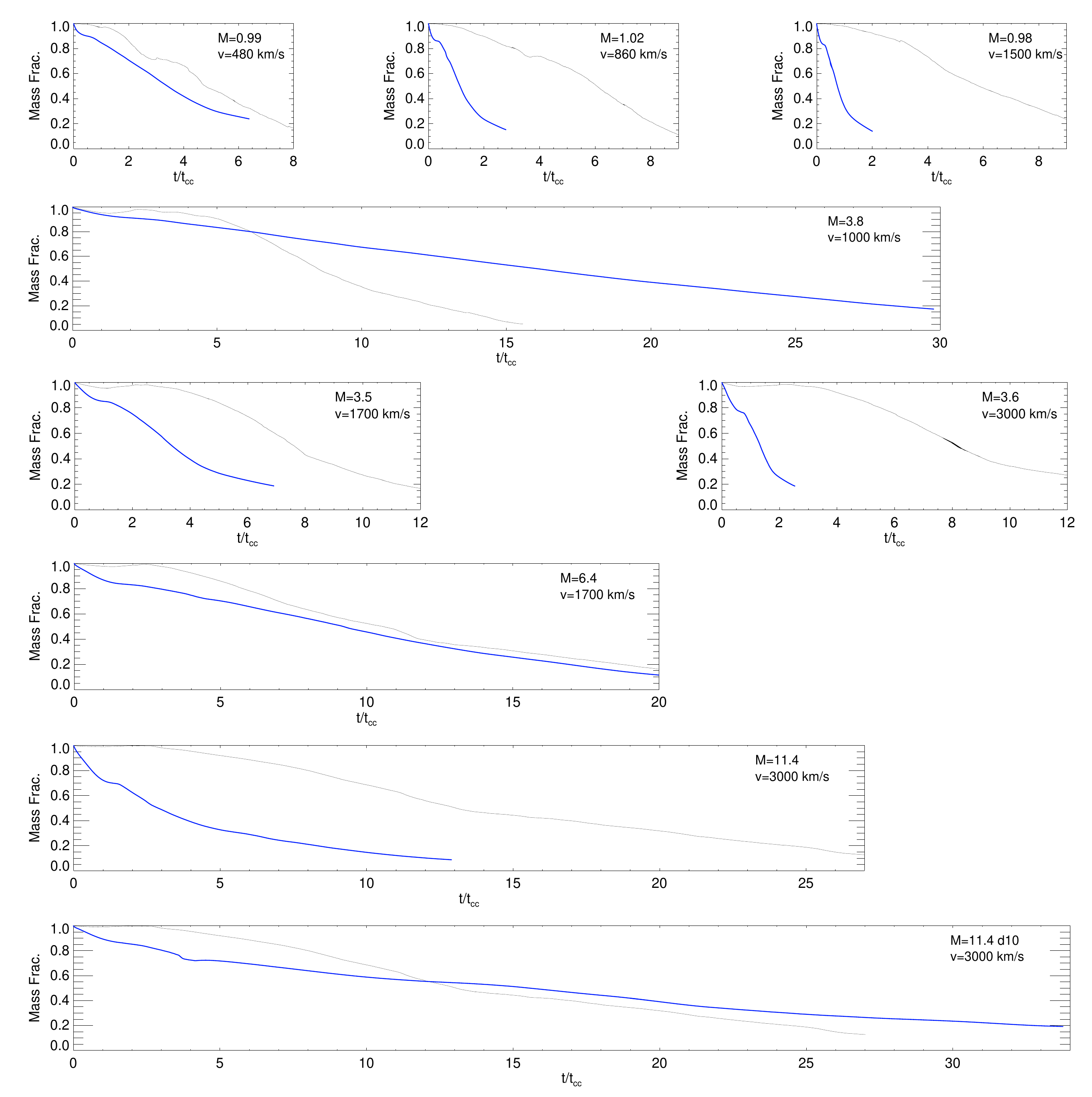}
\caption{Mass evolution for a selection of runs, labeled by Mach number and ambient velocity. Thick blue lines denote the conducting run from this paper and the thin black lines denote the results in the absence of thermal conduction, as taken from Paper I.}
\label{massplot}
\end{center}
\end{figure*}

In Fig.~\ref{massplot}, we show the mass retained by the cloud in each of these runs as a function of time in units of the cloud crushing time. More specifically, we plot $F_{1/3}(t)$, as defined in Paper I,  the fraction of total mass above 1/3 the original cloud density. Unlike for the cases without thermal conduction, the choice of the threshold fraction (1/3) does not affect the results since evaporative compression causes a strong separation between the hot and cold phases. Fig.~\ref{massplot} captures the full evolution of the 9 cases including the high density model, which is at the bottom of the plot. The blue lines denote the results from the runs in this paper and the black lines are the results from the corresponding runs without thermal conduction from Paper I. It is clear from this figure that the KH instability does not play a significant role in the mass evolution. As found in Orlando (2004) the mass loss is dominated by evaporation off the cold cloud and not by hydrodynamic instabilities. Moreover, we see that in a few instances the lifetime of the clouds is longer when thermal conduction is switched on - contrary to what one might expect. The reasons lie in the balance of conduction and cooling as we will see below.

\begin{figure*}
\begin{center}
\includegraphics[trim=0 0 0 0, width=1.0\textwidth]{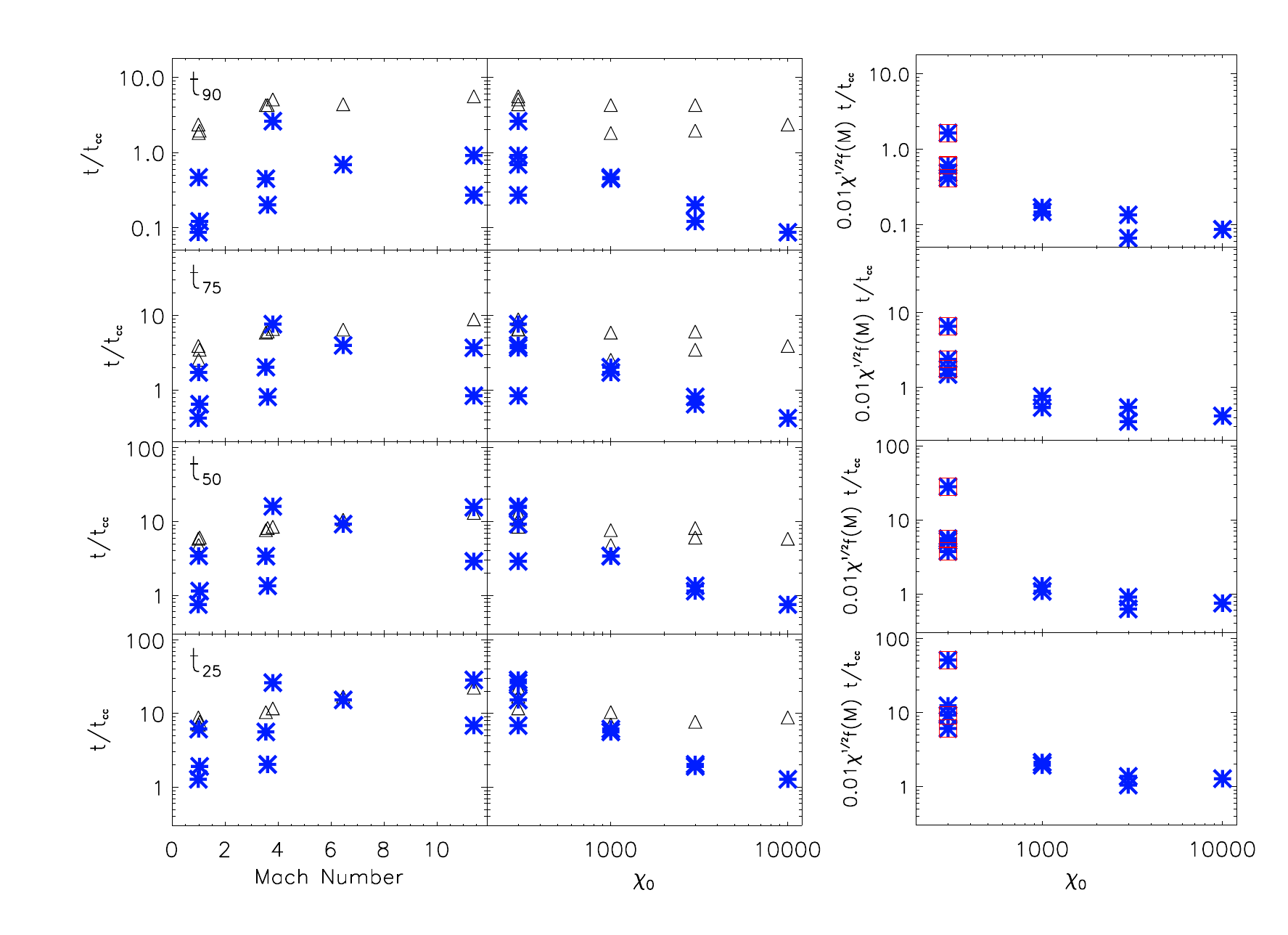}
\caption{Left column: Time for cloud to reach 90\%, 75\%, 50\%, and 25\% of its original mass as a function of the Mach number, $M$, of the ambient flow. Triangles are from previous runs not including the effect of conduction. Center column: Time for cloud to reach 90\%, 75\%, 50\%, and 25\% of its original mass  as a function of the initial density contrast, $\chi_0$, of the ambient flow. Right column:  $0.01 t/\tcc \times \chi_0^{1/2} f(M)$ as a function of  $\chi_0$. The boxed points relate to the outliers: $\chi 300v1000$, $\chi 300v3000$ and $\chi 300v1700$.}
\label{timeplot}
\end{center}
\end{figure*}

In Paper I, we were able to show that in the case without thermal conduction the lifetime of the cloud is almost completely dependent on Mach number.  However, when we plot the disruption times of conducting clouds as a function of the Mach number, no clear trend emerges, as shown in the leftmost column of Fig.~\ref{timeplot}.  On the other hand,  when we plot the disruption times as a function of $\chi_0$ as in the central panels of this figure, we see that they scale roughly as $\chi_0^{-1/2}$. 

In order to explain this dependence, we consider the balance between the internal energy that impinges on the cloud and the energy evaporated off it. The rate with which energy impinges on a cloud with cross-section $\pi R_{\rm c}^2$ is given by
\begin{equation}
\dot{E}_{\rm heat} = \frac{3\cps^2}{2\gamma} \rho_{\rm h} v_{\rm h}\pi R_{\rm c}^2 ,
\end{equation}
where $\cps$ is the post-shock sound speed, $v_{\rm h}$ is the velocity of the hot medium, $\rho_{\rm h}$ is the density of the hot medium and $\gamma$ is the ratio of specific heats. This heat goes into evaporating material off the cloud to a temperature beyond the peak of the cooling curve that allows it to decouple from the cloud. This temperature can be determined by measuring the mean temperature in the simulation within a density bracket of $[10^{-26},10^{-27}]$ g cm$^{-3}$. We find that this temperature, $T_{\rm evap}$, is largely uniform across runs at  $\approx 3 \times 10^6$ K, a value just beyond the peak of $\Lambda(T)$, where radiative cooling becomes much less efficient.
Note that the energy of the evaporating material is dominated by thermal energy, which is $\approx 3$ higher than its kinetic energy. 

If we require this rate to balance the heating rate modulo a constant factor $\eta_{\rm h}$, we get
\begin{equation}
{1\over 2} \dot{m} c_{\rm evap}^2 = - \eta_{\rm h} \frac{3\cps^2}{2\gamma} \rho_{\rm h} v_{\rm h}\pi R_{\rm c}^2 , \label{eq:mdot1}
\end{equation}
where $c_{\rm evap}^2 = \gamma k T_{\rm evap}/(\mu m_{\rm p})$, and 
the post-shock sound speed is given by
\begin{align}
&\cps^2 =\frac{\gamma k T_{\rm ps}}{\mu m_{\rm p}} =  \frac{\gamma k (10^4 \rm K)}{\mu m_{\rm p}}  \chi_0 \nonumber \\
&\max\left\{1, \frac{[(\gamma-1)M^2+2][2\gamma M^2-(\gamma-1)]}{4 (\gamma+1)^2M^2} \right\}  \equiv c_{\rm c}^2 \chi_0 f(M),
\label{eq:f(M)}
\end{align}
where $c_{\rm c}^2 \equiv \gamma k_{\rm B} 10^4 {\rm K}/{\mu m_{\rm p}} =$ 15 km/s  is the sound speed inside the cloud (compare Eq. 17 and 50 in \cite{1989ApJ...336..979B}; see also \cite{1988ApJ...326..769L}). Note that $\cps^2$ is 1/4 of what it would be without conduction because of the modification of the shock jump conditions by thermal conduction, as described in Appendix A.

Dividing both sides of Eq.~(\ref{eq:mdot1}) by the initial mass of the cloud, $m=\rho_{\rm c} 4 \pi R_{\rm c}^3/3$, and expressing the mass loss rate in terms of the cloud crushing time, $\tcc$, we obtain
\begin{equation}
\dot{\tilde{m}}\equiv \frac{\dot m}{m}\tcc 
 = - A \chi_0^{1/2} f(M) , 
\label{eq:mdot3}
\end{equation}
where  
\begin{equation}
A \equiv \eta_{\rm h} \frac{9}{4 \gamma} \frac{10^4 {\rm K}}{T_{\rm evap}} .
\end{equation}
In the rightmost column of Fig.~\ref{timeplot} we plot $t/\tcc \times \dot{\tilde{m}}$ as given by Eq.~(\ref{eq:mdot3}), which shows most points reasonably level except for three outliers.  Here we find that the value $A=0.01$ yields a good fit for most data points. If we assume that $T_{\rm evap} = 3 \times 10^6$ K, this implies that $\eta_{\rm h} = 2$. Alternatively, if we want to keep $\eta_{\rm h} = 1$, this would imply $T_{\rm evap} = 1.5 \times 10^6$ K.

\begin{figure*}[t]
\begin{center}
\includegraphics[trim=0 0 0 0, width=0.9\textwidth]{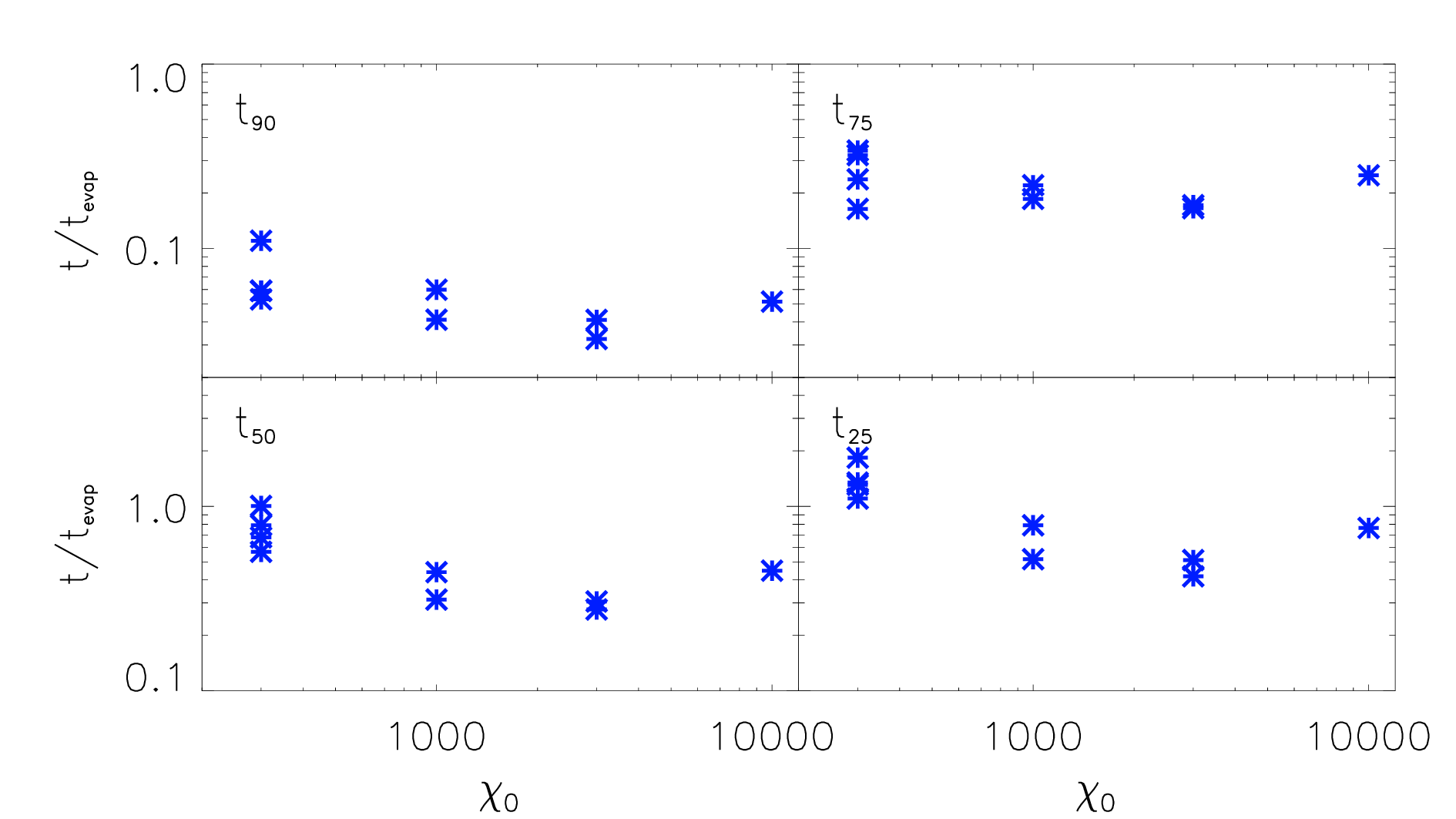}
\caption{Timescale for evaporation scaled as $t/t_{\rm evap}$ as in Eq.~(\ref{eq:tevapscale}) as a function of the initial density ratio $\chi_0$ for $t_{\rm 90}$, $t_{\rm 75}$, $t_{\rm 50}$ and $t_{\rm 25}$.}
\label{chi_factor2}
\end{center}
\end{figure*}

The outliers are the runs for which radiative cooling in the evaporative flow is significant. 
Including optically thin cooling inside the blob, Eq.~(\ref{eq:mdot1}) becomes

\begin{align}
&{1\over 2} \dot{m} c_{\rm evap}^2 = - \eta_{\rm h} \frac{3\cps^2}{2\gamma} \rho_{\rm h} v_{\rm h}\pi R_0^2 \nonumber \\
&+\eta_{\rm c} (4 \pi R_0^3/3) \left [ \frac{\dot{m}}{\pi R_0^2\cps}\right ]^2 \frac{\Lambda}{(\mu m_{\rm p})^2} ,
\end{align}
where again $\eta_{\rm c}$ is a constant factor. This can be rewritten as

\begin{align}
&\frac{\dot{m}}{m} = -\eta_{\rm h} \frac{9}{4\gamma} \frac{v_{\rm h}}{\chi_0 R_0} \frac{\cps^2}{c_{\rm evap}^2} \nonumber \\
&+\eta_{\rm c} (4 \pi R_0^3/3)^2 \frac{2\rho_{\rm c}}{\pi^2R_0^4\cps^2c_{\rm evap}^2} \frac{\Lambda}{(\mu m_{\rm p})^2}\left ( \frac{\dot{m}}{m}\right )^2 ,
\end{align}
which yields after expressing $\dot{m}$ in terms of $\dot{\tilde{m}}$

\begin{equation}
\dot{\tilde{m}} = - A \chi_0^{1/2} f(M) + B \dot{\tilde{m}}^2\frac{M}{\chi_0 f(M)} , \label{eq:mdot4}
\end{equation}
where 
\begin{equation}
B \equiv \eta_{\rm c} \frac{16 R_{\rm c} n_{\rm i,c}}{9c_{\rm evap}^2c_{\rm c}}\frac{\Lambda}{\mu m_{\rm p}}. \nonumber
\end{equation}
Note that $B$ is purely a function of the initial column density of the cloud, $R_{\rm c} \rho_{\rm c}.$
We can solve Eq.~(\ref{eq:mdot4}), choosing the negative root in the quadratic formula as we are looking for solutions where  $\dot{\tilde{m}}<0$.  This gives
\begin{equation}
\dot{\tilde{m}} = A f(M) \chi_0^{1/2} \frac{1-\sqrt{1+4g}}{2g}\, ,
\end{equation}
where  $g \equiv A\, B\, M \chi_0^{-1/2}$.
This implies that the timescale for evaporation, $t_{\rm evap}$ should be
\begin{equation}
\frac{t_{\rm evap}}{\tcc} \equiv \frac{-1}{\dot{\tilde{m}}} = \frac{1}{A f(M) \chi_0^{1/2}}\frac{2g}{\sqrt{1+4g}-1} . \label{eq:tevapscale}
\end{equation}
In Fig.~\ref{chi_factor2}, $t/t_{\rm evap}$ is plotted as a function of the density ratio $\chi_0$.  
Here we have taken $\eta_{\rm c} = 0.5$ and $\Lambda = 10^{-22}$ erg cm$^3$/s appropriate for the
the $\approx 3 \times 10^6$ evaporating gas, which has been heated beyond the peak of the cooling function.
Plugging in constants, one gets
\begin{align}
&g =  3.5 \Lambda_{-22}\left (\frac{\eta_{\rm c}}{0.5} \right )\left (\frac{A}{0.01} \right ) \nonumber \\
&\left (\frac{n_{\rm c} R_{\rm c}}{ 3 \times 10^{20} {\rm cm}^{-2}} \right )
\left (\frac{3 \times 10^6 K}{ T_{\rm evap}} \right ) M \left(\frac{1000}{\chi_0} \right)^{1/2}. \label{eq:g}
\end{align}
When the cooling term is insignificant, then $g$ is small and that means that
$\sqrt{1+4g} \approx 1+2g$ such that $(\sqrt{1+4g}-1)/2g \approx 1$. When the cooling term is large, then $(\sqrt{1+4g}-1)/(2g) \approx 1/\sqrt{g}$. Interestingly, 
apart from the decreased temperature jump across the shock, none of this depends on the conduction law itself.

\section{Morphology}

\begin{figure*}
\centerline{\includegraphics[trim=0 0 0 0,width=0.995\textwidth]{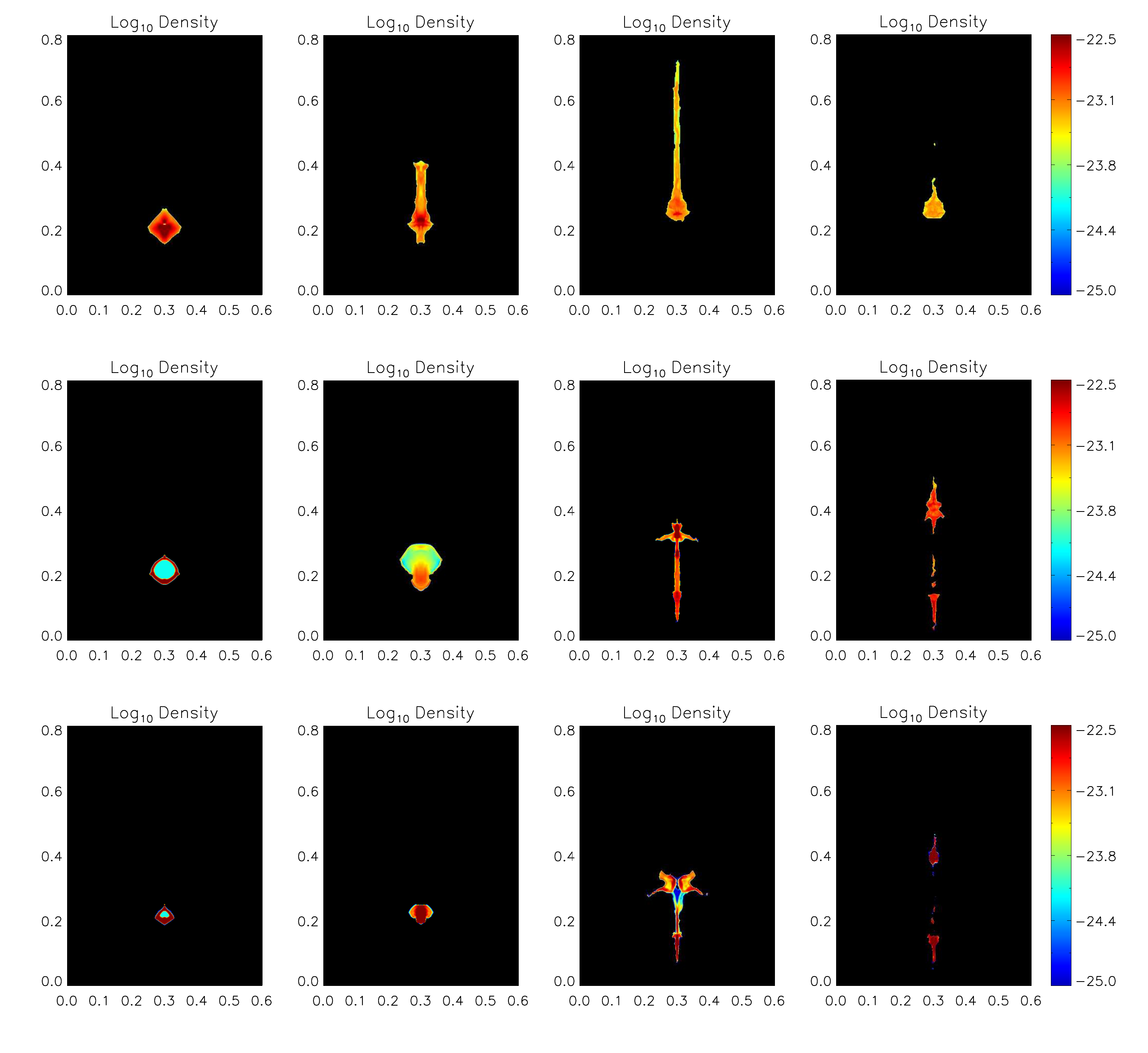}}
\caption{Density (log) slices from simulations with Mach numbers $\approx 1$, at times at which  the fraction of the mass at or above 1/3 the original density of the cloud is 90\% ($t_{\rm 90},$ first column), 75\% ($t_{\rm 75},$ second column), 50\% ($t_{\rm 50},$ third column), and 25\% ($t_{\rm 25},$ fourth column).  {\em First row:}  $\chi 1000v480$  {\em Second row:}  $\chi 3000v860$ {\em Third row:} $\chi 10000v1500.$  All lengths are given in kpc, with $R_{\rm c} = 0.1$ kpc, and the density range has been chosen to emphasize the evolution of the compressed cloud material.}
\label{fig:M1d}
\end{figure*}

\begin{figure*}
\centerline{\includegraphics[trim=0 0 0 0,width=0.995\textwidth]{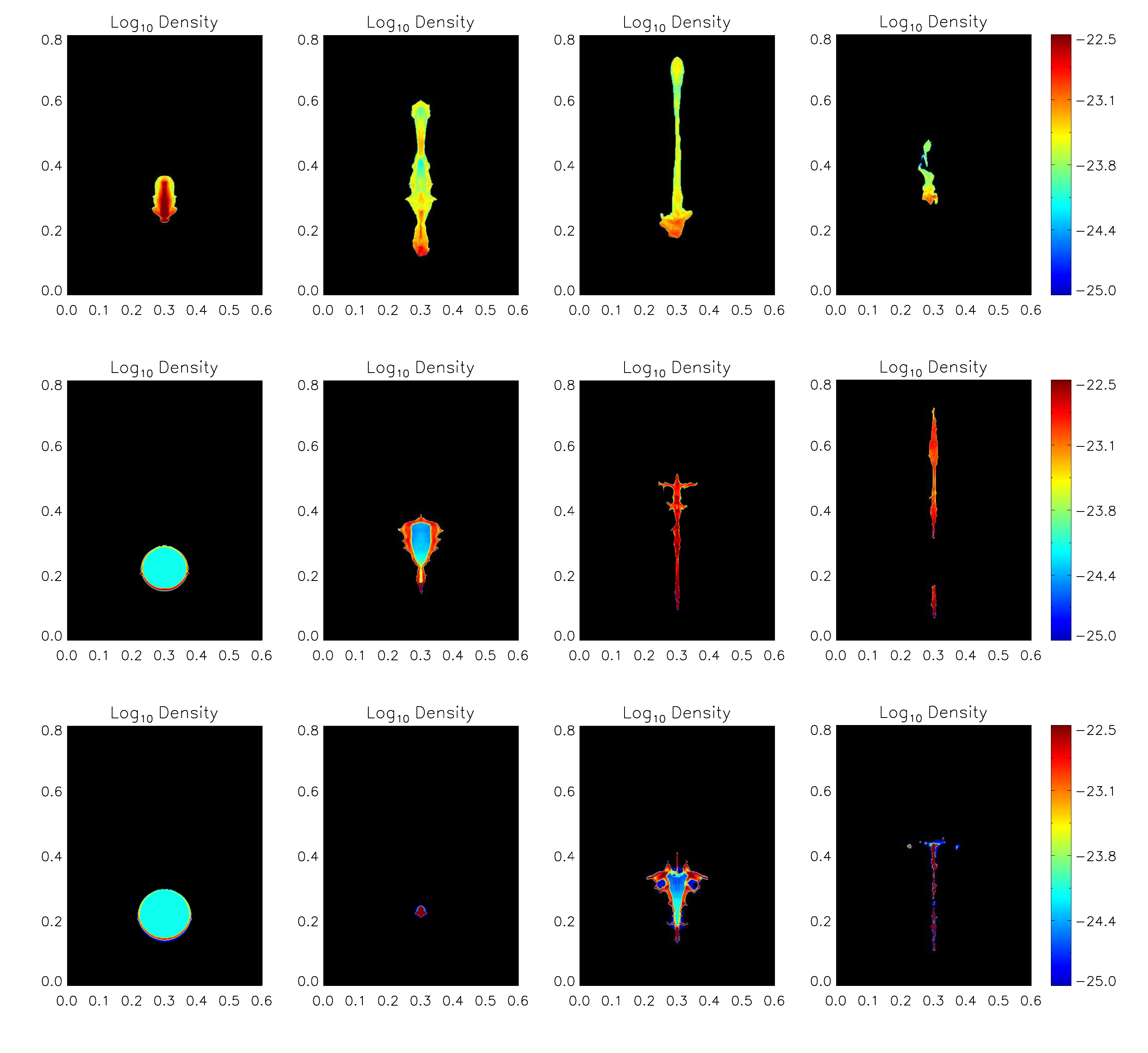}}
\caption{Density (log) slices from simulations with Mach numbers $\approx 3$, at $t_{\rm 90}$ (first column), $t_{\rm 75},$ (second column), $t_{\rm 50}$ (third column), and $t_{\rm 25}$, (fourth column).  {\em First row:}  $\chi 300v1000$ {\em Second row:} $\chi 1000v1700$ {\em Third row:} $\chi 3000v3000.$ All lengths are given in kpc,  with $R_{\rm c} = 0.1$ kpc.}
\label{fig:M3d}
\end{figure*}

\begin{figure*}
\centerline{\includegraphics[trim=0 0 0 0,width=0.995\textwidth]{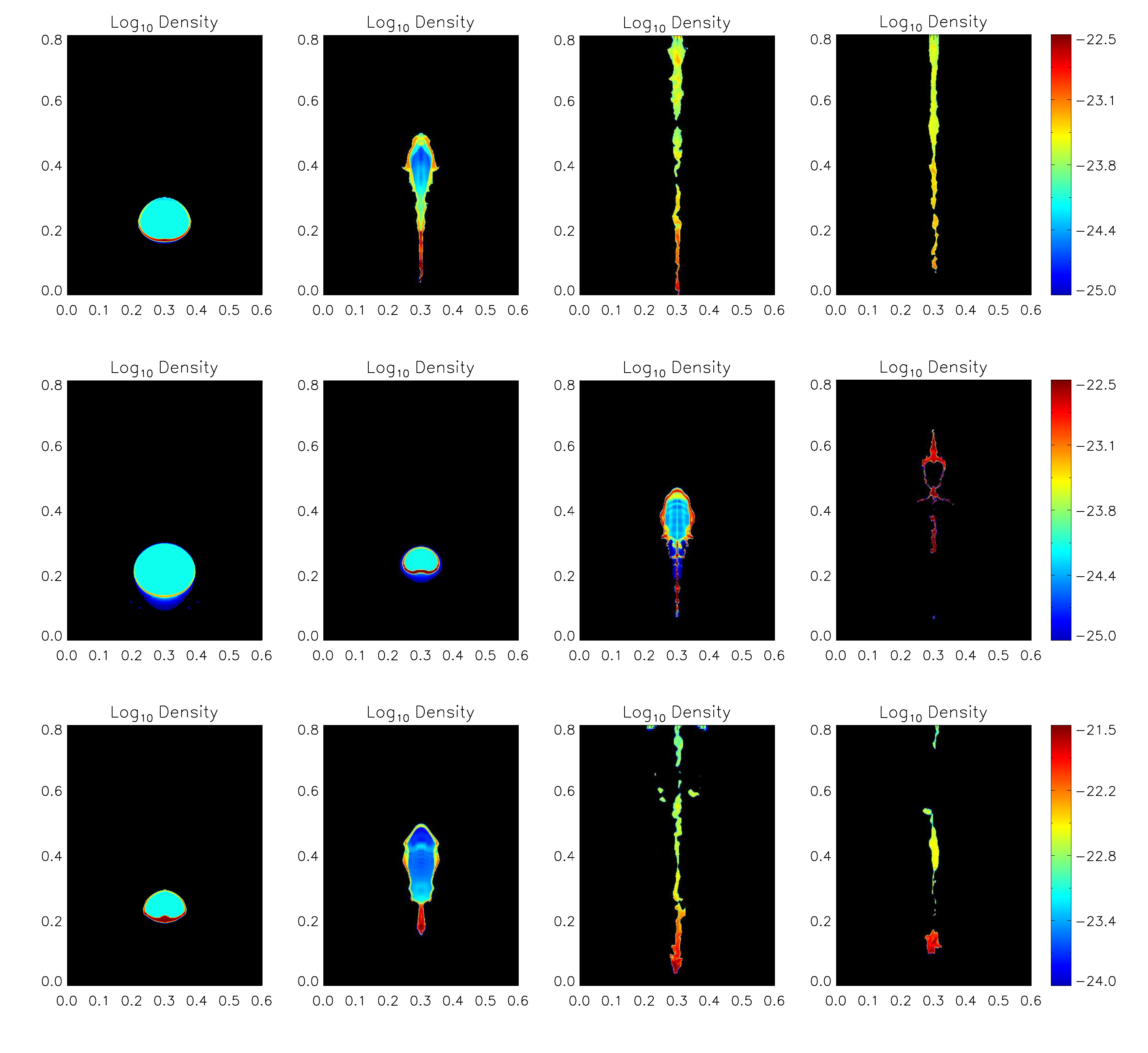}}
\caption{Density (log) slices from simulations with Mach numbers $\approx 6$ and $\approx 11$, at at $t_{\rm 90}$ (first column), $t_{\rm 75},$ (second column), $t_{\rm 50}$ (third column), and $t_{\rm 25}$, (fourth column).  {\em First row:} $\chi 300v1700$  {\em Second row:} $\chi 300v3000$ {\em Third row:} $\chi 300v3000$ at ten times cloud density. All lengths are given in kpc,  with $R_{\rm c} = 0.1$ kpc.}
\label{fig:M11d}
\end{figure*}

\begin{figure*}
\centerline{\includegraphics[width=1\textwidth]{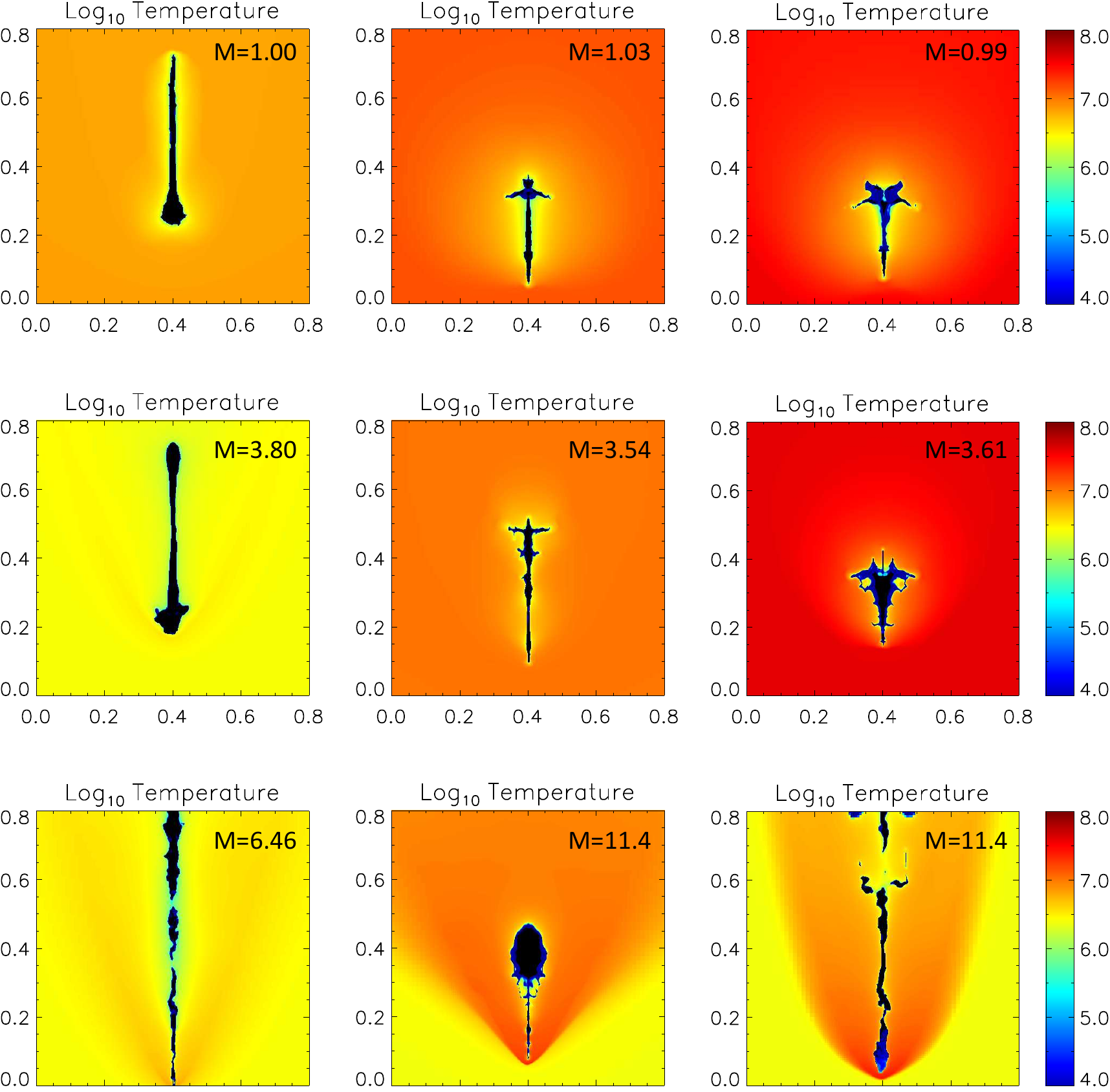}}
%\vspace{0.2in}
\caption{Temperature (log) slices from all simulations at time $t_{\rm 50}$ with increasing $M$ from top left to bottom right.
In order these are 
$\chi 1000v480,$  $\chi 3000v860,$ $\chi 10000v1500$ (Top row), 
$\chi 300v1000,$ $\chi 1000v1700,$ $\chi 3000v3000$ (Second row),  
$\chi 300v1700$  $\chi 300v3000,$ and $\chi 300v3000$ at ten times cloud density (Third row). All lengths are given in kpc, with $R_{\rm c} = 0.1$ kpc.}
\label{fig:Temp}
\end{figure*}

To better illustrate the evolution of the clouds, in Figs.~\ref{fig:M1d}-\ref{fig:M11d} we plot slices of the central density distribution at $t_{\rm 90}$, $t_{\rm 75}$,  $t_{\rm 50}$  and $t_{\rm 25}$.  While the morphological evolution is relatively similar across the runs, the timescale of that evolution is not.
At $t_{\rm 90},$ the clouds shrink and get denser as they are compressed by the evaporative flow. Interestingly, this compression is the highest in the cases with the lowest Mach number and lowest values of $\chi_0.$   By setting the pressure behind the shock moving into the cloud due to evaporation equal to $p_{\rm evap} = - \dot m c_{\rm evap}/ (4 \pi R_{\rm c}^2),$ we can estimate the timescale for the cloud compression as 
\be
t_{\rm comp} \approx \frac{R_{\rm c}}{\sqrt{\gamma p_{\rm evap} /\rho_c}} = \frac{R_{\rm c}}{\sqrt{\gamma R_{\rm c} c_{\rm evap}/ (3 t_{\rm evap})}}.
\ee
Again assuming that $T_{\rm evap} = 3 \times 10^6$ K this gives
\be
t_{\rm comp} \approx t_{\rm evap} \left[ \frac{M}{30} \frac{\tcc}{t_{\rm evap}}\right]^{1/2}.
\ee
This means that compression is most significant, i.e. that $t_{\rm comp}/t_{\rm evap}$ is smallest, when $M$ is small and the evaporation time is long compared to $\tcc$.
Thus, for the lowest $\chi_0$ value cases at $M \approx 1$ and $M\approx 3,$ the exterior pressure has managed to compress the entire cloud by $t_{\rm 90}$, while for the $M\approx 1$ cases with larger $\chi_0$ values the radius of the cloud is roughly halved.   On the other hand, at higher $M$ and $\chi_0$ values,  only the  outer shell of the cloud is compressed and the density of the core of the cloud is lower. 

At $t_{\rm 75},$ the clouds have a range of morphologies, depending on the ratio between $t_{\rm comp}$ and  $t_{\rm evap}.$ For the low and moderate $M$ and $\chi_0$ cases, the clouds have moved past the stage of initial compression and have begun to be stretched into filaments that are extremely narrow in the direction perpendicular to the flow, but extended in the streamwise direction.  As in the case without conduction (see Paper I), the clouds are stretched mostly by the streamwise pressure gradient, rather than by shearing from the ambient medium.  In these simulations, the density structure of the developing filament is complex: a lower density core is encased in a higher density shell, and the back of the cloud is flared due to the slightly lower pressure on the downstream side of the cloud.  The flow around the filament is fairly laminar and there is little turbulent mixing. This can be quantified in terms of the dimensionless Peclet number, Pe, that is defined as the ratio between the advection rate and the diffusion rate, or $L v \rho c_{\rm p}/\kappa$, where $L$ is a typical length scale, $v$ is the velocity, $\rho$ is the density, $c_{\rm p}$ is the specific heat at constant pressure and $\kappa$ is the thermal conductivity. The flow outside of the very sharp boundary in density between the filament and the ambient flow has low Peclet numbers ($<10^{-2}$) which means that turbulent transport is negligible compared to the evaporative transport.

On the other hand, in the runs with larger $M$ and $\chi_0$ values, the clouds only reach the initial stage of compression at $t_{75},$ and they have not yet begun to be stretched at this time.   In the fiducial $\chi300v3000$ run with $M=11.4$, the cloud is still in the midst of its initial compression, but due to additional cooling, the $\chi300v3000$ run  with ten times the initial density (and a larger $t_{\rm evap}$)  is much further in its evolution.   The difference between these two $M=11.4$ runs highlights the importance of column density in the evolution of the cloud,  not only in determining the minimum cloud size as in eq.\ (\ref{eq:heatcool}), but also in  determining the rate of radiative cooling in the evaporating material for clouds above this threshold.

By $t_{\rm 50},$ even the fiducial $M=11.4$  run with relatively efficient cooling has produced a filament that has reached a length of at least twice the original cloud diameter.  At each Mach number, the  filaments flare at the downstream end, except in the cases  in which  the most compression times have passed and the interface between the cloud material and the ambient medium has become smoother. Furthermore, in the $\chi1000v480$ ($M=1.0$) and $\chi300v1000$ ($M=3.8$) cases, the leading edge of the filament has developed a bulbous morphology, trailed by a long cometary tail. Finally, at $t_{\rm 25},$ evaporative compression becomes significant in all cases, the filaments become very thin, and the clouds split into chains of smaller clumps that will soon be indistinguishable from the surrounding medium.  

In  Fig.~\ref{fig:Temp} we show slices of the temperatures  at $t_{\rm 50}$ for all the runs. The filamentary clouds are all cold with $T<10^4$ K material surrounded by a  small corona that then blends into the hot ambient medium.  As discussed in the Appendix, the temperature jumps seen in the $M \geq 4$ runs occur just downstream from the shocks themselves, with important consequences for the conductive flux and the postshock temperature.

\section{Velocity Evolution}

\begin{figure*}
\begin{center}
\includegraphics[trim=0 0 0 0, width=1.0\textwidth]{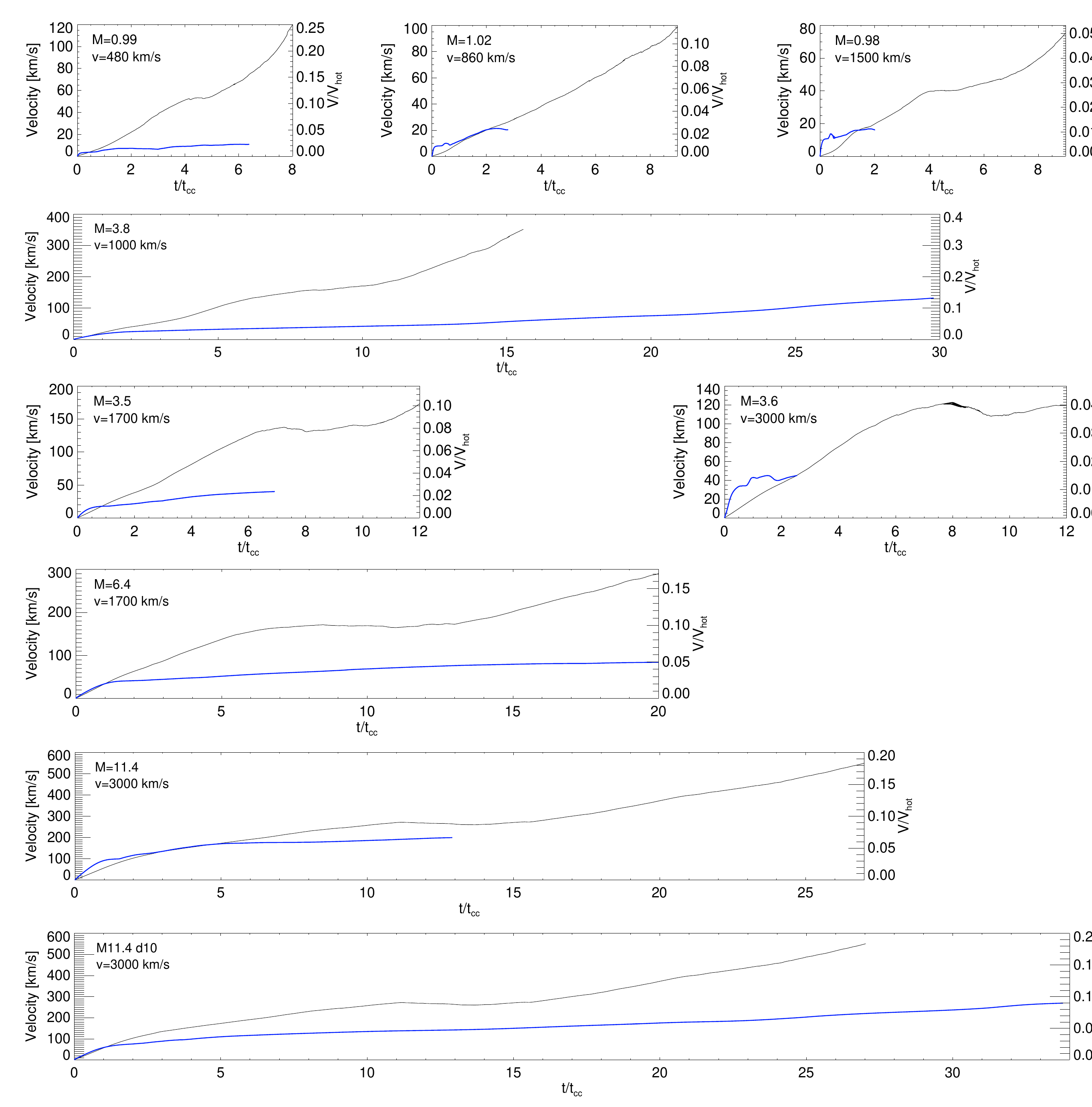}
\caption{Velocity evolution for a selection of runs,  labeled by Mach number and ambient velocity. Thick blue lines denote the conducting run from this paper and the thin black lines denote the results in the absence of thermal conduction.}
\label{velplot}
\end{center}
\end{figure*}

The impact of electron thermal conduction on cold cloud acceleration is even more dramatic than on the cloud disruption described above. In Fig.~\ref{velplot}, we show the velocity evolution from our conduction runs, contrasted with the results without conduction from Paper I for the same choices of velocity and Mach number.   Here we see that at  early times the velocity of the clouds exceeds that in the runs without conduction, but soon afterwards the acceleration of the cold clouds becomes extremely inefficient.  This is because in all our conducting runs, the cloud either evaporates  within $\approx 2 t_{\rm cc}$ and is no longer distinguishable from the ambient medium or it compresses into a thin, long filament that has a small cross-section to the ambient wind. In the absence of conduction, the cloud velocity is set by the momentum imparted by the hot wind, which depends on the cross-section that the cloud presents to the hot flow. In Paper I, we estimated this velocity as
\be
v_{\rm estimate}(t)  = \frac{3 v_{\rm h} t}{4 \chi_0^{1/2} t_{\rm cc}}. 
\label{eq:vest}
\ee
At early times, this estimate describes some of the runs with thermal conduction relatively accurately. However, from Fig.~\ref{velplot}, it is apparent that in all cases the velocity attained by the cloud at late times, remains a small fraction of this value. 

The large differences in the velocity evolution of the clouds are due to two important effects.  The first is the compression of the clouds by the evaporating flow, which then causes it to present a very small cross-section to the wind after a time $\approx t_{\rm comp},$
strongly limiting further acceleration.  In addition, the evaporation of the cloud is biased in the downstream direction, yielding an extra increase in velocity $\propto  (-\dot{m}/m)\, c_{\rm evap}$ due to the rocket effect, similar to what occurs in photoevaporative flows  \cite[e.g,][]{1955ApJ...121....6O,1990ApJ...354..529B}.   This is the cause of the increase in the cloud velocity at early times with respect to the results from Paper I.

\begin{figure*}
\begin{center}
\includegraphics[trim=0 0 0 0, width=1.0\textwidth]{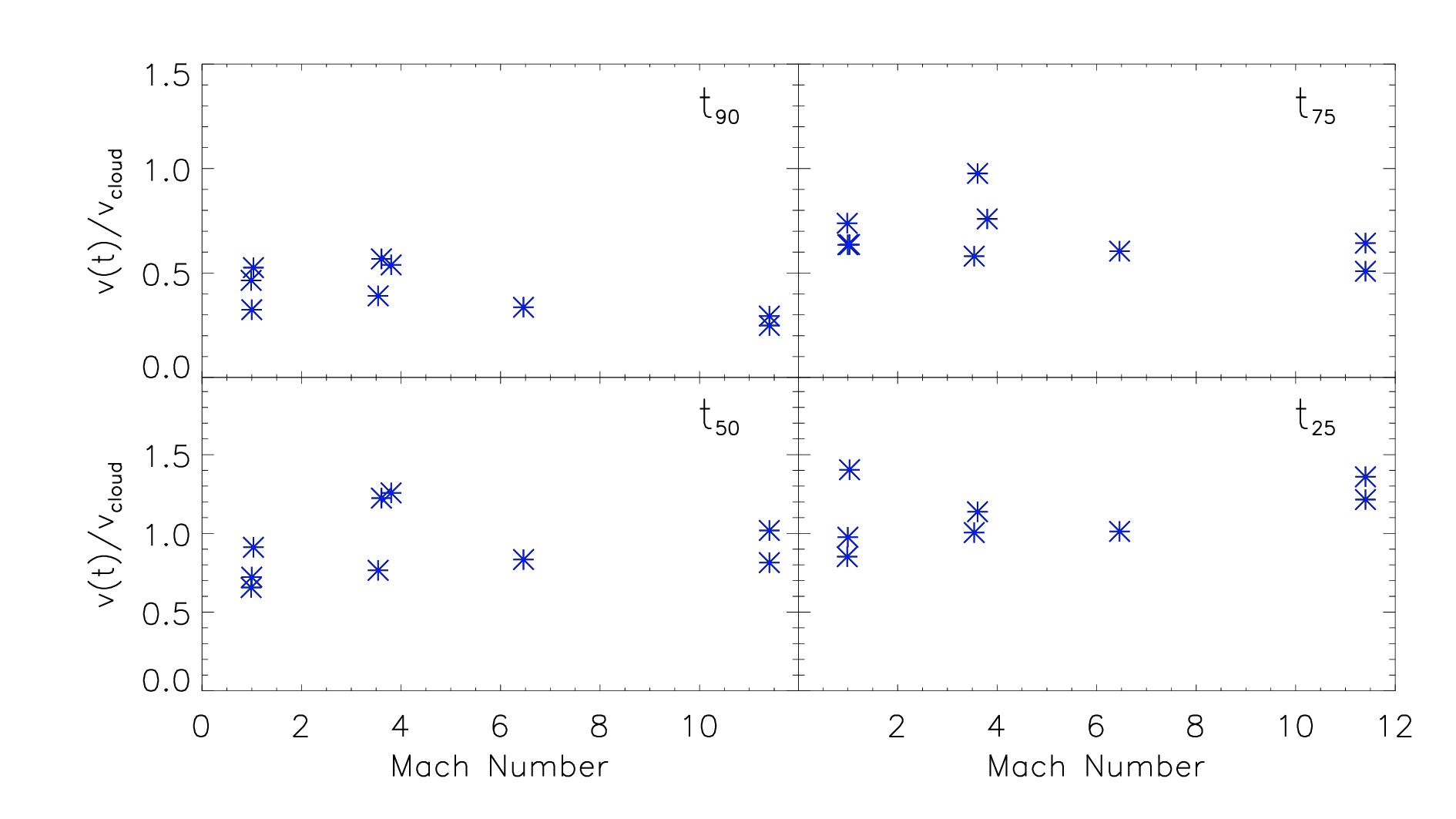}
%v_chi.pro
\caption{Velocity of cloud as a function of Mach number scaled by $v_{\rm c}$ as given by  Eq.~(\ref{eq:vcloud}).}
\label{velplot2}
\end{center}
\end{figure*}

Adding this additional source of acceleration to Eq.\ (\ref{eq:vest}), rewriting $-\dot{m}/m$ as $1/t_{\rm evap}$, and 
plugging in the compression time for $t$, we get
\be
v_{\rm c} \approx \eta_{\rm shock} \frac{v_{\rm h}}{\chi_0^{1/2}} \left[ \frac{M}{30} \frac{t_{\rm evap}}{\tcc}\right]^{1/2} + \eta_{\rm rocket} \, c_{\rm evap} \left[ \frac{M}{30} \frac{t_{\rm cc}}{t_{\rm evap}}\right]^{1/2},
\label{eq:vcloud}
\ee
where $\eta_{\rm shock}$ and $\eta_{\rm rocket}$ are geometrical factors that quantify the efficiency of acceleration by the impinging flow and the evaporating material, respectively.   In Fig.~\ref{velplot2}, we have plotted $v/v_{\rm c}$ as given by Eq.~(\ref{eq:vcloud}) with $\eta_{\rm shock}=0.4$ and $\eta_{\rm rocket}=0.6,$ which gives a reasonable fit over our entire range of simulations.  Note that in the cases in which cooling is negligible ($M$ small and $\chi_0$ large), $t_{\rm evap}/\tcc \propto \chi_0^{-1/2}$ and the evaporation term is dominant, 
while in cases in which cooling is significant, $t_{\rm evap}$ is much longer and the shock term is most important.

In all cases, however, acceleration is inefficient with the cloud velocity reaching at most $15\%$ of $v_{\rm h}$ at late times.   Thus,
in order for cold material entrained by a hot wind to reach velocities of the order of several hundred km/s as seen in some star-forming galaxies \citep{2005ApJS..160..115R,2006asup.book..337M} either other accelerating forces must be acting on the cloud or thermal conduction must be strongly suppressed by magnetic fields.

\section{Dependence on Numerical Resolution}

\begin{figure*}
\begin{center}
\includegraphics[trim=0 0 0 0, width=0.49\textwidth]{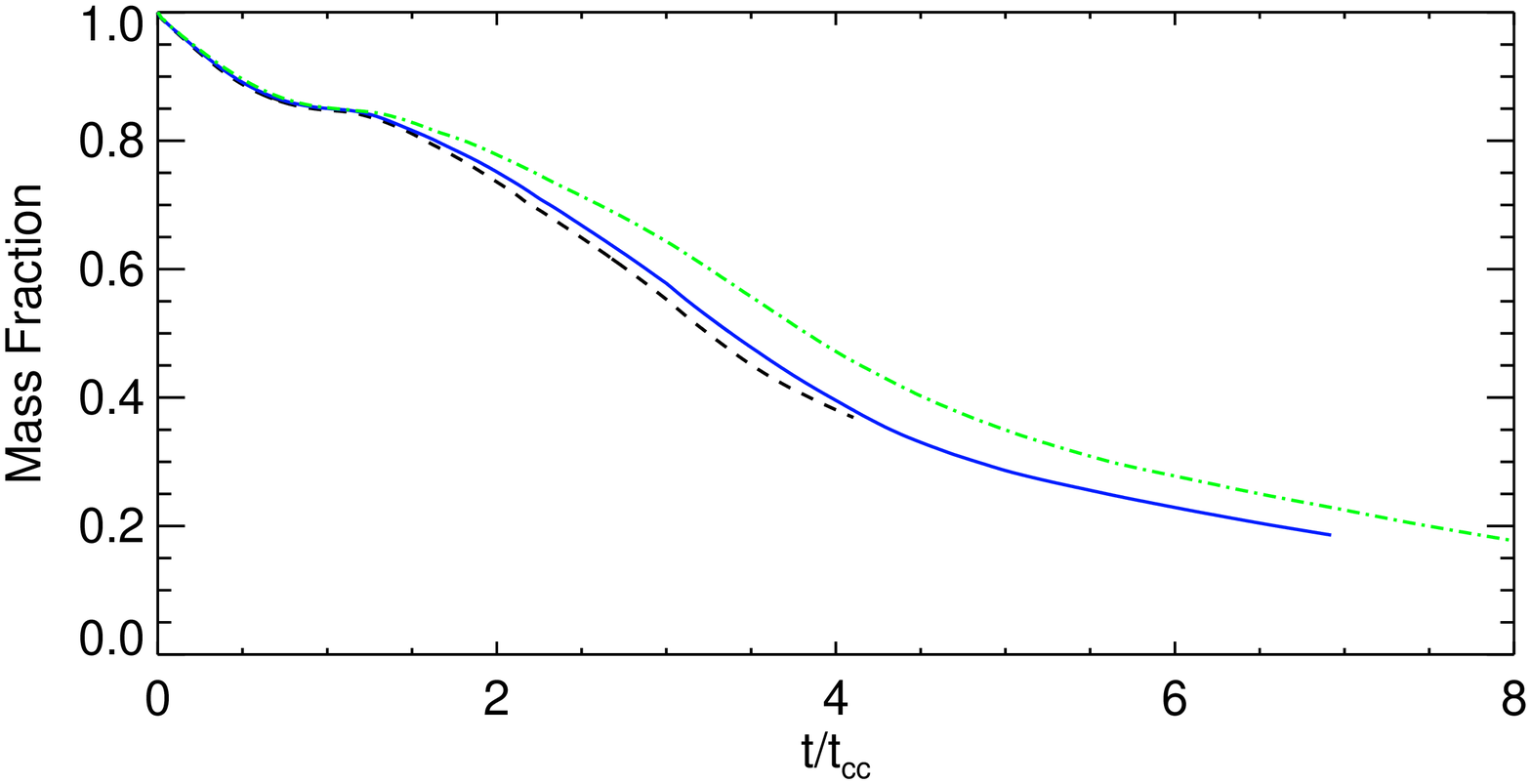}
\includegraphics[trim=0 0 0 0, width=0.49\textwidth]{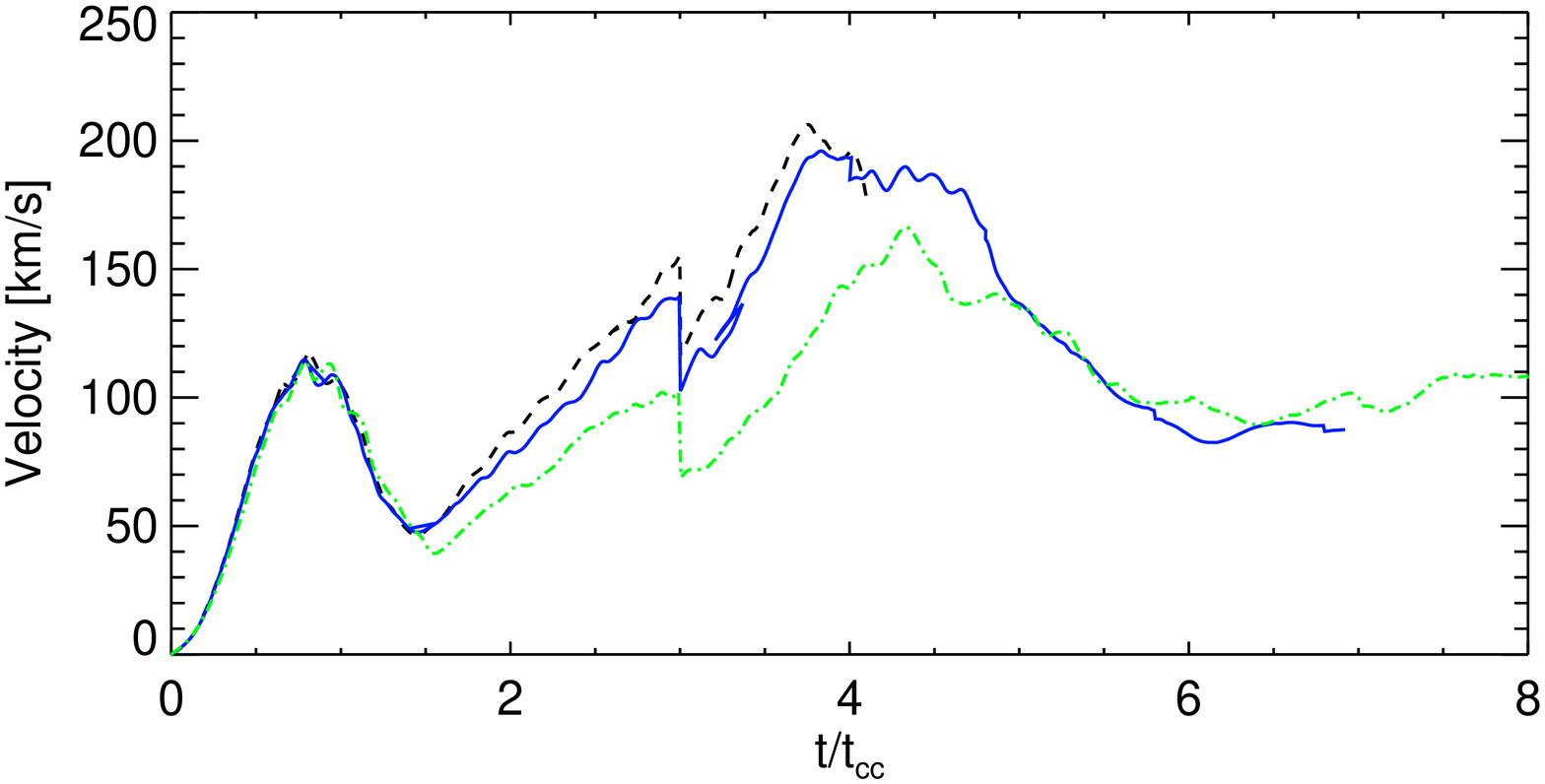}
\caption{Mass evolution, $F_{1/3}(t)$, (left) and cloud velocity (right) for run $\chi 1000v1700$ with three different grid resolutions: the fiducial resolution at $\Delta x =4.8\times 10^{18}$ cm (blue - solid),  $\Delta x =3.2\times 10^{18}$ cm (black - dashed) and $\Delta x =9.6\times 10^{18}$ cm (green - dash-dotted).}
\label{convergence}
\end{center}
\end{figure*}

Finally, we turn our attention to numerical convergence, which can  particularly be a concern for problems that involve shearing instabilities and radiative cooling. Several authors recommend that the cloud radius be resolved with $\approx 100$ cells \citep{1994ApJ...434L..33M, 2009ApJ...698..693F}.
Furthermore \cite{2010ApJ...722..412Y} pointed out that in order to capture the Rayleigh-Taylor and Kelvin-Helmholtz instabilities when radiative cooling is included, one must resolve the cooling length $l_{\rm cool}\equiv v_{\rm h} t_{\rm cool}=3kT_{\rm c} v_{\rm h}/\Lambda \approx 4\times10^{18}\, {\rm cm}(v_{1000})\Lambda_{-22}^{-1}.$  However, in our case, thermal conduction leads to a transition layer between the cold cloud and the hot, ambient medium which is thicker than $l_{\rm cool}$ and which damps these instabilities. Thus resolving $l_{\rm cool}$ is not expected to be necessary.

A second important length scale is the Field length, $\lambda_{\rm F}=\sqrt{(\kappa(T) T/(n^2 \Lambda)}$ \citep{1965ApJ...142..531F, 1990ApJ...358..375B}, the maximum length scale across which thermal conduction can dominate over radiative cooling. For Spitzer conduction  $\lambda_{\rm F} \approx 1.3 \times 10^{20} T_7^{7/4}n^{-1} \Lambda_{-22}^{-1/2}$ cm, which implies that in all our runs, in the ambient medium $\Delta x \ll \lambda_{\rm F}$ and inside the cloud $\Delta x \gg \lambda_{\rm F}$. However, inside the cloud our setup does not allow any cooling because at temperatures $T<10^4$ K, cooling is switched off. As soon as material evaporates, the high pressure will not allow for fragmentation due to cooling, and thus resolving the Field length is not likely to be an issue.

To confirm these expectations, we have run our fiducial simulation $\chi1000v1700$ with 4 (instead of 5) levels of refinement corresponding to a cell size of  $\Delta x=9.6\times 10^{18}\, {\rm cm}=R_{\rm c}/32$ and also with a finer resolution than our fiducial run. In this case, the computing time requirements for a run with 6 levels of refinement were too high, so we settled for a run with a 50\% higher peak resolution by increasing the number of blocks in each direction by 50\%, yielding a cell size of $\Delta x =3.2\times 10^{18}\, {\rm cm}=R_{\rm c}/96$. Since we enforce maximal resolution in and around the cold cloud, this minimum cell size was always maintained in the crucial parts of the simulation volume. This higher-resolution run took roughly 200,000 core hours on the Stampede supercomputer. 

In Fig.~\ref{convergence}, we have plotted the evolution of the cloud mass, $F_{1/3}$, and cloud velocity as a function of time in these runs. The resulting curves for $F_{1/3}$ agree well, especially between the fiducial and the high-resolution run. Also the cloud velocities agree reasonably well between the fiducial and the high-resolution run. However, the low-resolution run with $\Delta x =9.6\times 10^{18}$ cm has a slightly higher mass and a resulting lower cloud velocity than the more highly resolved runs - a result of a non-negligible numerical viscosity.  This implies that the spatial resolution chosen for our fiducial runs is sufficient and the scalings found in this paper do not depend on the grid size of the simulation.

\section{Discussion and Summary}

We have performed a suite of AMR hydrodynamical simulations of the evolution cold clouds in a hot wind, including both radiative cooling and electron thermal conduction.   Observations of such $\approx 10^4$ K clouds are the primary probe of high-redshift galaxy outflows, and their evolution is highly dependent on the cloud  column density, $n_{\rm i,c} R_{\rm c}.$  If $n_{\rm i,c} R_{\rm c} <  1.3 \times 10^{18} {\rm cm}^{-2} T_7^2,$ the electron mean-free path is larger than the cloud radius, leading to runaway heating that quickly disrupts the cloud.  On the other hand, if $n_{\rm i,c} R_{\rm c}$ is above this threshold, cloud evolution is much more complex and gradual, and we have focused on this more interesting high-column density case in this study.

In this case, we have carried out nine different runs that span the range of conditions found in outflowing starburst galaxies, plus two more at varying spatial resolutions.   As the equations that govern the evolution of the cloud are invariant under any transformation in which ${\bf x} \rightarrow \alpha {\bf x},$ ${\bf t} \rightarrow \alpha {\bf t}$ {\rm and} $\rho \rightarrow \rho/\alpha,$ these simulations also allow us to draw conclusions over a wide range of cloud properties.  In all cases, the presence of electron thermal conduction drastically changes  the morphological evolution, mass loss rate, and velocity evolution of the interacting clouds. While the outer layers of the clouds evaporate owing to thermal conduction, the cores are stretched into dense and cold filaments, as was also found by \cite{2002A&A...395L..13M}, \cite{2004ApJ...604...74F}, \cite{2005A&A...444..505O}, and \cite{2013ApJ...766...45J}. The simulations show how a conductive interface forms, leading to an increased density in the core that manages to radiate the energy conducted into the cloud, while stabilizing the cloud against hydrodynamical instabilities. 

In Tab.~\ref{tab:disruption}, we list the times at which the masses of the clouds reach 90\%, 75\%, 50\% and 25\% of their original values, expressed in units of the cloud crushing time $t_{\rm cc} \equiv R_{\rm c} \chi_0/v_{\rm h}$, with  $\chi_0$ the initial density ratio between the cloud and the ambient medium, which is moving at a velocity $v_{\rm h}$.  While in the absence of conduction the disruption time scales as $t_{\rm cc} \sqrt{1+M}$, when thermal conduction is taken into account, the disruption times instead scale roughly as $t_{\rm cc}/\sqrt{\chi_0}$.   We show that this dependence can be explained by considering the balance between the internal energy that impinges on a spherical cloud and the energy evaporated off the cloud.  In particular, the heat impinging on the cloud goes into heating the material in the outer layers to a temperature beyond the peak of the cooling curve, which allows it to decouple from the cloud and evaporate.

In fact, accounting for radiative cooling in the evaporative flow yields an even better description of our results:
\begin{equation}
\frac{t_{\rm evap}}{\tcc}  \approx \frac{100}{f(M) \chi_0^{1/2}}\frac{2g}{\sqrt{1+4g}-1} ,
\end{equation}
where $g$ is given by Eq.~(\ref{eq:g}) and $f(M)$ by Eq.~(\ref{eq:f(M)}). 
Thus, we can capture the combined effects of conduction and radiative cooling with a fairly simple analytic form, as shown in Fig.~\ref{chi_factor2}. Interestingly, this scaling does not depend on the form of the conduction law itself, apart from its role in affecting the temperature jump across the bow shock ahead of the cloud.   In the downstream region the temperature is about 1/4 of the value that we would find for a shock in the absence of thermal conduction, and in the appendix of the paper we should how this value is determined by the electron sound speed and the fact that the temperature front moves out in front of the density discontinuity.

By considering the force of the evaporating flow on the cloud, we can also estimate the timescale for the cloud compression to be 
\be
t_{\rm comp} \approx t_{\rm evap} \left[ \frac{M}{30} \frac{\tcc}{t_{\rm evap}}\right]^{1/2},
\ee
implying that at the point at which the same fraction of the material is evaporated, compression will be most significant when the Mach number is small and when radiative losses in the evaporative flow are significant, extending $t_{\rm evap}.$   These trends are in good correspondence to the morphologies observed in our simulations, which
evolve from compression in an outer shell, to collapse to dense core, to stretching out into a dense filament that extends in the streamwise direction.

This evolution means that the cloud only presents an appreciable cross section to the wind for a time $\approx t_{\rm comp}.$  As a result the acceleration of the cold clouds by the ambient medium becomes  inefficient, even though the presence of an evaporative flow has the potential to increase the velocity of the cloud due to the rocket effect. Estimating that the cloud acceleration occurs only on a timescale $\approx t_{\rm comp}$ yields
\be
v_{\rm c} \approx \eta_{\rm shock} \frac{v_{\rm h}}{\chi_0^{1/2}} \left[ \frac{M}{30} \frac{t_{\rm evap}}{\tcc}\right]^{1/2} + \eta_{\rm rocket} \, c_{\rm evap} \left[ \frac{M}{30} \frac{t_{\rm cc}}{t_{\rm evap}}\right]^{1/2},
\label{eq:vcloudsecond}
\ee
where $\eta_{\rm shock}$ and $\eta_{\rm rocket}$ are geometrical factors that quantify the efficiency of acceleration by the impinging flow and evaporated material.  Again this is a good description of our numerical results tabulated in Tab.~\ref{tab:disruption},  with $\eta_{\rm shock}=0.4$ and $\eta_{\rm rocket}=0.6,$ giving the best match.  We should add one caveat about our assumption to prevent cooling below temperatures of $10^4$ K while several observations of outflows do show a molecular component \citep[e.g.][]{2014Natur.516...68G}. With an HI photo-ionization cross-section of $6.3 \times 10^{-18}$ cm$^2$, the cloud simulated in this paper would become optically thick with a neutral fraction of around 0.1\%. Hence, for high enough column densities, shielding from the background might become important and lead to further cooling. This could result in longer cloud lifetimes and should be investigated using a full radiative transfer scheme.\\

Our simulations appear robust against changes in the resolution of the computational grid. In a convergence test, we found that the cloud mass does not change significantly between the $\Delta x =R_{\rm c}/32$ and $\Delta x =R_{\rm c}/96$ runs, and the velocity evolution does not change significantly between $\Delta x =R_{\rm c}/64$ $\Delta x =R_{\rm c}/96.$  Finally, one can verify which regime of conduction applies to our setup. The energy flux that must pass through a radius $R$ around the cloud is $F_{\rm evap} \approx 0.5 \dot{m} c_{\rm evap}^2/4 \pi R^2$ and the density around the cloud is $\dot{m}/4 \pi R^2 / c_{\rm evap}$. This means that the ratio of the conductive flux into the cloud divided by the saturated flux is $F_{\rm evap}/F_{\rm sat} =  0.5 c_{\rm evap}^3 / (5 \phi c_{\rm s,iso}^3)  = 0.1/\phi (c_{\rm evap}/c_{\rm s,iso})^3$. Since the sound speed in the evaporating medium is approximately $c_{\rm evap}$, this ratio is much less than 1, and the conductive flux is unsaturated. The condition for efficient conduction is when $\phi$ is sufficiently big for the saturated conductive heat flux to move faster than the energy flux of material flowing at its sound speed.  Or in other words, energy needs to be able to move upstream in a $M \approx 1$ flow for conductive evaporation to operate efficiently.  If magnetic fields suppress this, the evaporation times found above could increase significantly.

If the suppression of conduction by magnetic fields is more modest, however, the mass and velocity scalings given above can be applied with significant confidence to future numerical and observational studies. In galaxy-scale simulations in which the evolution of individual clouds cannot be modeled with the required resolution directly, such clouds could be easily added as subgrid tracer particles, following our scalings for mass loss and cloud velocity.  Observationally, our results can be used to help determine the possible origins of  clouds given their velocities and positions with respect to the outflowing galaxy.  A  general conclusion in this case is that the acceleration of the cold clouds by the ambient medium is very inefficient in the presence of thermal conduction.

This means one of two things.   Either magnetic fields must both strongly suppress conduction, or  cold clouds with outflow velocities several 100 km/s cannot have been accelerated from $v=0$ purely by hot, outflowing material.  The estimates of $F_{\rm eval}/F_{\rm sat}$ above suggest that conduction must be suppressed by a magnetic fields by a factor of at least $\approx 10$ to prevent cloud evaporation.  In fact, at larger $\phi,$ values the only place in which the conduction rate explicitly determines our results is in setting the value of the post-shock temperature, and decreasing conduction would increase this temperature not decrease it.  If conduction indeed operates in this regime, then the most promising model for the origin of the cold clouds may be precipitation directly from the hot wind as a result of radiative cooling \citep[e.g.][]{1995ApJ...444..590W,2000MNRAS.317..697E,2003ApJ...590..791S,2011ApJ...740...75W,2015ApJ...803....6M,2016MNRAS.455.1830T}, rather than being swept up by the wind.  On the other hand, in the case without conduction, \cite{2015MNRAS.449....2M} find that clouds with tangled internal magnetic fields are difficult to disrupt and the cloud fragments end up moving with the same speed as the external medium.   Further investigation on the effects of the magnetic fields on the evolution of cold clouds in supersonic flows is clearly needed.

\acknowledgements
We thank the referee for many helpful comments on the manuscript, and thank Romeel Dav\'e, Paul E. Dimotakis,  Timothy Heckman, Crystal Martin, Eve Ostriker, Eliot Quataert, and Todd Thompson, for helpful discussions. ES gratefully acknowledge the Simons Foundation for funding the workshop {\em Galactic Winds: Beyond Phenomenology} which helped to inspire this work.  We would also like to acknowledge the Texas Advanced Computing Center (TACC) at The University of  Texas at Austin, and  the Extreme Science and Engineering Discovery Environment (XSEDE) for providing HPC resources via grants TG-AST130021 and TG-AST140004 that have contributed to the research results reported within this paper.  Part of the simulations were run on the JUROPA/JURECA supercomputers at the Juelich Centre for Supercomputing under grant 9059. The FLASH code was developed in part by the DOE-supported  Alliances Center for Astrophysical Thermonuclear Flashes (ASC) at the University of Chicago. ES was supported by NSF grant AST11-03608 and NASA theory grant NNX15AK82G.

\section*{Appendix A: Shock Jump Conditions with Conduction}

Here we examine how the temperature jump at the incoming shock  is modified by saturated thermal conduction.   Approximating the shock as planar and perpendicular to the direction of the flow, the equations of mass, momentum and energy conservation are
\ba
\rho_1 v_1 & = & \rho_2 v_2 ,  \\
P_1+\rho_1 v_1^2 & = & P_2+\rho_2 v_2^2,  \\
\rho_1v_1 \left( \frac{1}{2}v_1^2 \right) & = & \rho_2 v_2\left( \frac{1}{2}v_2^2 +\frac{\gamma}{\gamma-1} \frac{P_2}{\rho_2}\right) +F_2,
\ea
where in the last equation $F_2 \equiv 5\phi\ P_2^{3/2}\rho_2^{-1/2}$ and 
we have neglected the upstream thermal pressure and its conductive flux. Comparing to the conduction law assumed in this paper, Eq.\ (\ref{eq:thermalcond}), $5\phi p^{3/2}\rho^{-1/2}=0.34 c_{\rm s,e}n_{\rm e}k_{\rm B} T$ and assuming $n_{\rm e}/n_{\rm total}=0.507$, $\mu= 0.6$ yields $\phi \approx 1.1$. Here the subscripts 1 and 2 denote upstream and downstream quantities, respectively. Radiative losses are not important for the shock structure in the conditions covered in this paper. 

\begin{figure*}[t]
\begin{center}
\includegraphics[trim=0 0 0 0, width=0.48\textwidth]{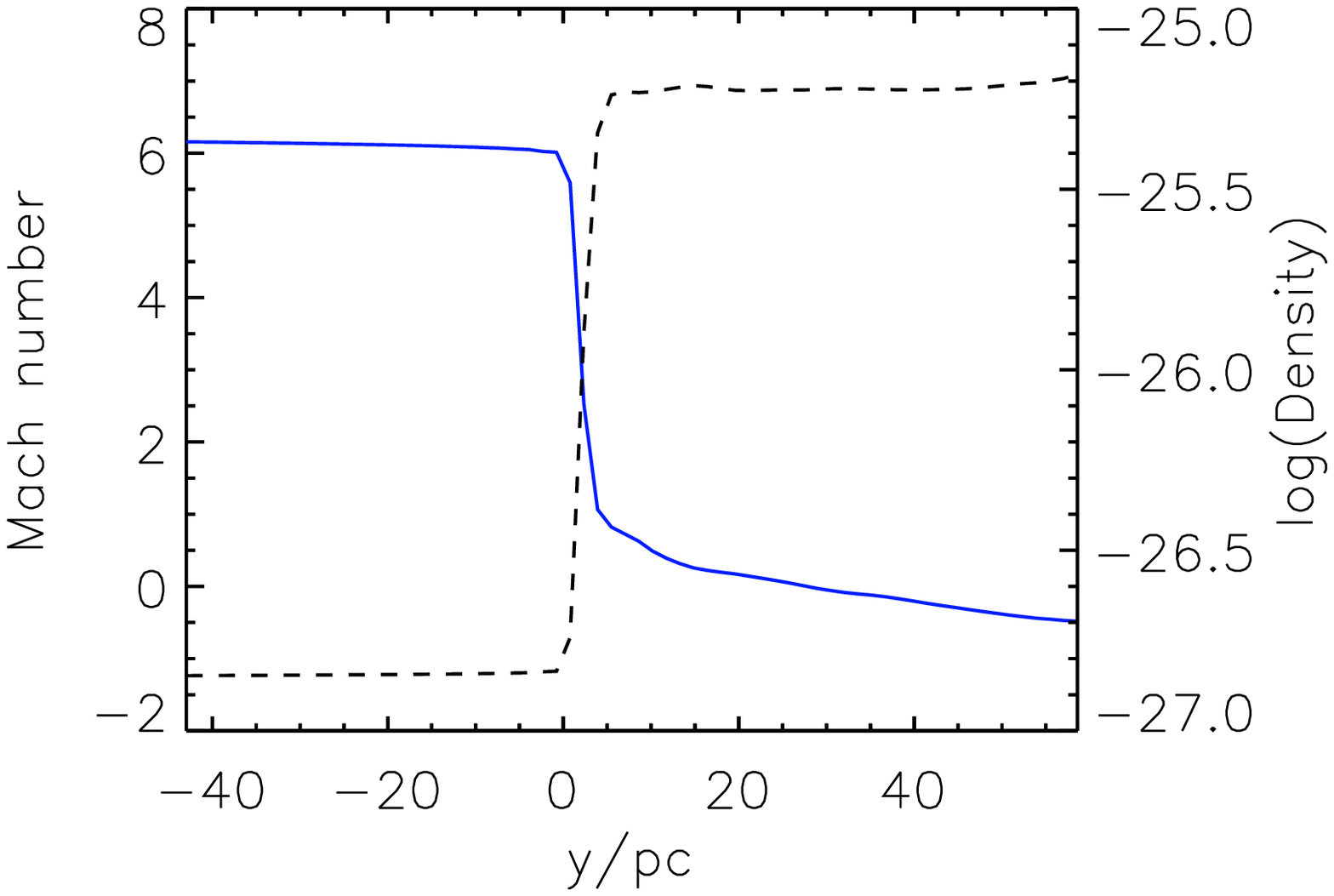}
\includegraphics[trim=0 0 0 0, width=0.48\textwidth]{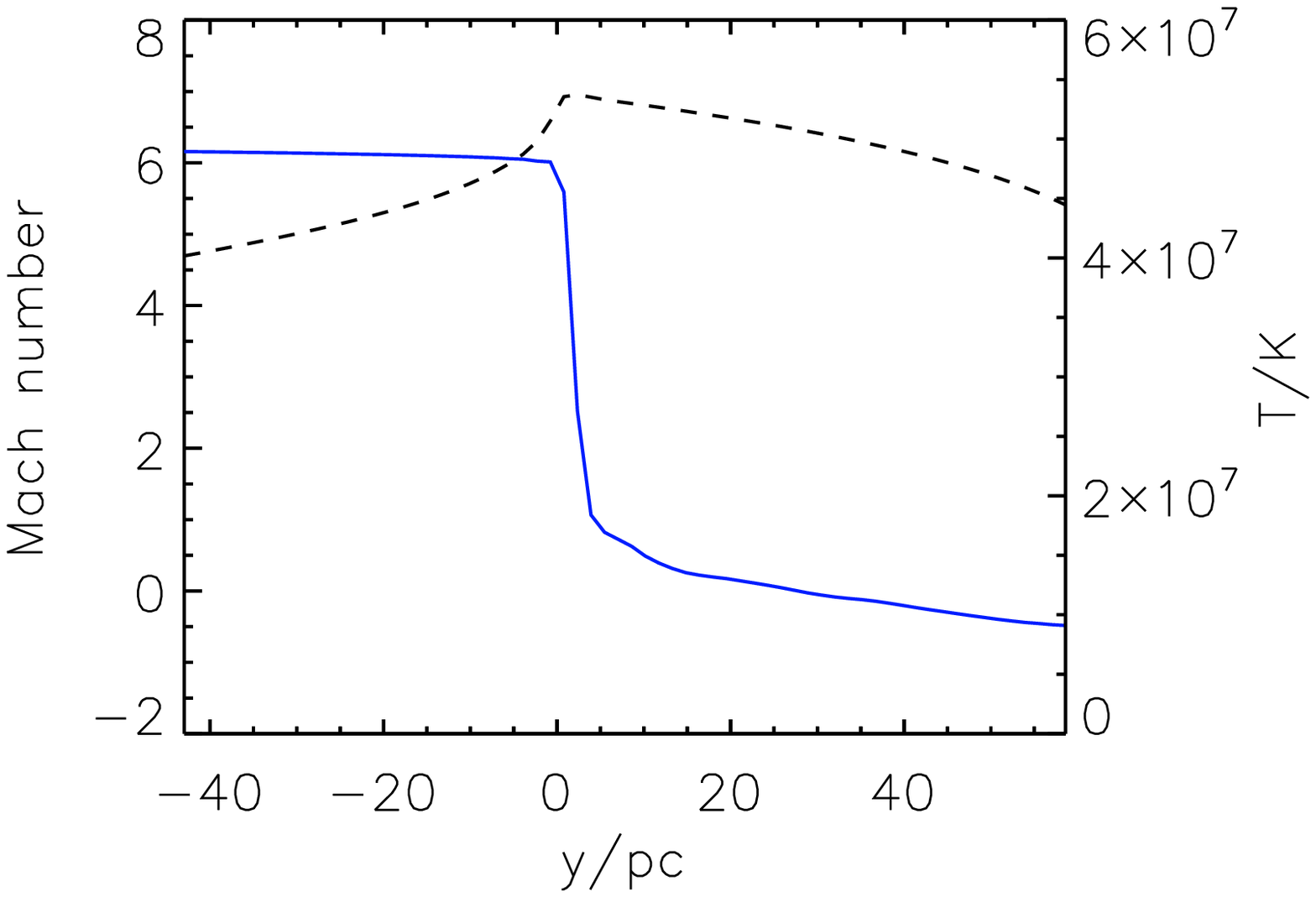}
\caption{Density (left) and temperature (right) jumps across the bow shock for run $v_{\rm h}=1000$ km/s and $\chi_0=300$ at $t=\tcc$. The upstream Mach number $v_{\rm h}/c_{\rm h}$ is shown in both plots as a blue, solid line.}
\label{fig:appendix_1}
\end{center}
\end{figure*}

From mass and momentum conservation we can derive the jump condition for the velocity as
\be
\frac{P_2}{\rho_2} = v_2 \left [  v_1(1+1/M_{\rm iso}^2) -v_2\right ] ,
\ee
where $M_{\rm iso}\equiv v_1(\rho_1/P_1)^{1/2}$ is the isothermal Mach number. 
After a bit of algebra, we can also compute the ratio of velocities as
\be
\frac{v_2}{v_1} = \frac{5}{8} + \frac{5}{4M_{\rm iso}} -\sqrt{ \left( \frac{5}{8} + \frac{5}{4M_{\rm iso}} \right )^2 +\frac{F_2}{2\rho_1 v_1^3} -\frac{1}{4}- \frac{5}{4M_{\rm iso}}} . \label{eq:app_v1}
\ee
In the absence of conduction and in the limit of high $M_{\rm iso}$, $v_2/v_1 = 1/4,$ and
$P_2/\rho_2 = 3v_1^2/16$. 
When the conductive term is nonzero, we can rewrite it as 
\be
\frac{F_2}{2\rho_1 v_1^3}= \frac{5\phi P_2^{3/2}}{2\rho_2^{3/2}v_1^2v_2} = \frac{5\phi}{2} \left (\frac{P_2}{\rho_2v_1v_2}\right )^{3/2} \left (\frac{v_2}{v_1}\right)^{1/2} .
\ee
Defining $x\equiv v_2/v_1$ and $y \equiv P_2({v_1v_2\rho_2})^{-1}=1+1/M_{\rm iso}^2-x,$ Eq.~(\ref{eq:app_v1}) can then be rewritten as 
\begin{align}
&x^2-\left (\frac{5}{4} + \frac{5}{2M_{\rm iso}}  \right)x - \frac{5\phi}{2} \left (1+\frac{1}{M_{\rm iso}^2}-x \right )^{3/2} x^{1/2} \nonumber \\
&+ \frac{1}{4}\left (1+\frac{5}{4M_{\rm iso}} \right )=0 . \label{eq:app_x}
\end{align}

We can solve this equation numerically as function of $\phi$ and $M_{\rm iso}.$ When we look at the temperature across the bow shock in our simulation, we see that the temperature front is leading the density discontinuity and hence sits at a location where the density is 1/4 of the postshock density (see Fig.~\ref{fig:appendix_1}).  This means that the conductive flux across the shock is set by the upstream density and the saturated flux is proportional to $\rho$ at a fixed temperature, i.e. $F_2=5\phi P_2^{3/2}/\rho_2^{1/2} \times (\rho_1/\rho_2)$, yielding an effective $\phi_{\rm eff}\approx 1.1/4\approx 0.27$.

In addition, the solution to Eq.~(\ref{eq:app_x}) only depends  weakly  on $M_{\rm iso}$. Still, we can determine the effective Mach number at the shock by taking for the shock speed
\begin{eqnarray}
c_{\rm ps}^2 &=& c_{\rm up}^2/f(M) \nonumber \\
&=&c_{\rm up}^2  {\rm max}\left [1, \frac{[(\gamma-1)M^2+2](2\gamma M^2-(\gamma-1))}{4(\gamma+1)^2 M^2} \right ] .
\end{eqnarray}
The Mach number of our shock is given by

\be
\frac{v_{\rm h}}{c_{\rm ps}} = {\rm min} \left [M, M  \times \frac{2(\gamma+1) M}{\sqrt{[(\gamma-1)M^2+2][2\gamma M^2-(\gamma-1)]}}\right ] . \label{eq:app_voverc}
\ee
Such that, for $\gamma = 5/3$, the maximum of $v_{\rm h}/c_{\rm ps}$ is $\approx 3.5$.

Hence, plugging in $F_2=5\phi_{\rm eff} P_2^{3/2}/\rho_2^{1/2}$ with $\phi_{\rm eff} =0.27$ and $M_{\rm iso}= 3.5 \gamma^{1/2},$ we find from Eq. (\ref{eq:app_x}) $x\approx 0.07$ which corresponds to the density jump $1/x\approx 14$ that we see in Fig.~\ref{fig:appendix_1}. Hence $P_2/\rho_2=y x v_1^2 \approx 0.25 \times (3/16)  v_1^2$.  So, for $\phi \approx 1.1,$ the post-shock temperature is about a quarter of the value found in the absence of conduction.

\bibliographystyle{apj}
\bibliography{Conduction}

\begin{thebibliography}{}
\expandafter\ifx\csname natexlab\endcsname\relax\def\natexlab#1{#1}\fi

\bibitem[{{Arrigoni Battaia} {et~al.}(2015){Arrigoni Battaia}, {Hennawi},
  {Prochaska}, \& {Cantalupo}}]{2015ApJ...809..163A}
{Arrigoni Battaia}, F., {Hennawi}, J.~F., {Prochaska}, J.~X., \& {Cantalupo},
  S. 2015, \apj, 809, 163

\bibitem[{{Barai} {et~al.}(2013){Barai}, {Viel}, {Borgani}, {Tescari},
  {Tornatore}, {Dolag}, {Killedar}, {Monaco}, {D'Odorico}, \&
  {Cristiani}}]{2013MNRAS.430.3213B}
{Barai}, P., {Viel}, M., {Borgani}, S., {et~al.} 2013, \mnras, 430, 3213

\bibitem[{{Begelman} \& {McKee}(1990)}]{1990ApJ...358..375B}
{Begelman}, M.~C., \& {McKee}, C.~F. 1990, \apj, 358, 375

\bibitem[{{Benson} {et~al.}(2003){Benson}, {Bower}, {Frenk}, {Lacey}, {Baugh},
  \& {Cole}}]{Bens+03}
{Benson}, A.~J., {Bower}, R.~G., {Frenk}, C.~S., {et~al.} 2003, \apj, 599, 38

\bibitem[{{Bertoldi} \& {McKee}(1990)}]{1990ApJ...354..529B}
{Bertoldi}, F., \& {McKee}, C.~F. 1990, \apj, 354, 529

\bibitem[{{Borkowski} {et~al.}(1989){Borkowski}, {Shull}, \&
  {McKee}}]{1989ApJ...336..979B}
{Borkowski}, K.~J., {Shull}, J.~M., \& {McKee}, C.~F. 1989, \apj, 336, 979

\bibitem[{{Cen} \& {Chisari}(2011)}]{2011ApJ...731...11C}
{Cen}, R., \& {Chisari}, N.~E. 2011, \apj, 731, 11

\bibitem[{{Chandrasekhar}(1961)}]{1961hhs..book.....C}
{Chandrasekhar}, S. 1961, {Hydrodynamic and hydromagnetic stability,
  International Series of Monographs on Physics, Oxford: Clarendon}

\bibitem[{{Cowie} \& {McKee}(1977)}]{1977ApJ...211..135C}
{Cowie}, L.~L., \& {McKee}, C.~F. 1977, \apj, 211, 135

\bibitem[{{Creasey} {et~al.}(2015){Creasey}, {Theuns}, \&
  {Bower}}]{2015MNRAS.446.2125C}
{Creasey}, P., {Theuns}, T., \& {Bower}, R.~G. 2015, \mnras, 446, 2125

\bibitem[{{Dalton} \& {Balbus}(1993)}]{1993ApJ...404..625D}
{Dalton}, W.~W., \& {Balbus}, S.~A. 1993, \apj, 404, 625

\bibitem[{{Dav{\'e}} {et~al.}(2011){Dav{\'e}}, {Oppenheimer}, \&
  {Finlator}}]{2011MNRAS.415...11D}
{Dav{\'e}}, R., {Oppenheimer}, B.~D., \& {Finlator}, K. 2011, \mnras, 415, 11

\bibitem[{{Efstathiou}(2000)}]{2000MNRAS.317..697E}
{Efstathiou}, G. 2000, \mnras, 317, 697

\bibitem[{{Erb} {et~al.}(2012){Erb}, {Quider}, {Henry}, \&
  {Martin}}]{2012ApJ...759...26E}
{Erb}, D.~K., {Quider}, A.~M., {Henry}, A.~L., \& {Martin}, C.~L. 2012, \apj,
  759, 26

\bibitem[{{Ferland} {et~al.}(1998){Ferland}, {Korista}, {Verner}, {Ferguson},
  {Kingdon}, \& {Verner}}]{1998PASP..110..761F}
{Ferland}, G.~J., {Korista}, K.~T., {Verner}, D.~A., {et~al.} 1998, \pasp, 110,
  761

\bibitem[{{Ferrara} {et~al.}(2005){Ferrara}, {Scannapieco}, \&
  {Bergeron}}]{2005ApJ...634L..37F}
{Ferrara}, A., {Scannapieco}, E., \& {Bergeron}, J. 2005, \apjl, 634, L37

\bibitem[{{Field}(1965)}]{1965ApJ...142..531F}
{Field}, G.~B. 1965, \apj, 142, 531

\bibitem[{{Fragile} {et~al.}(2004){Fragile}, {Murray}, {Anninos}, \& {van
  Breugel}}]{2004ApJ...604...74F}
{Fragile}, P.~C., {Murray}, S.~D., {Anninos}, P., \& {van Breugel}, W. 2004,
  \apj, 604, 74

\bibitem[{{Fryxell} {et~al.}(2000){Fryxell}, {Olson}, {Ricker}, {Timmes},
  {Zingale}, {Lamb}, {MacNeice}, {Rosner}, {Truran}, \&
  {Tufo}}]{2000ApJS..131..273F}
{Fryxell}, B., {Olson}, K., {Ricker}, P., {et~al.} 2000, \apjs, 131, 273

\bibitem[{{Fujita} {et~al.}(2009){Fujita}, {Martin}, {Mac Low}, {New}, \&
  {Weaver}}]{2009ApJ...698..693F}
{Fujita}, A., {Martin}, C.~L., {Mac Low}, M.-M., {New}, K.~C.~B., \& {Weaver},
  R. 2009, \apj, 698, 693

\bibitem[{{Geach} {et~al.}(2014){Geach}, {Hickox}, {Diamond-Stanic}, {Krips},
  {Rudnick}, {Tremonti}, {Sell}, {Coil}, \& {Moustakas}}]{2014Natur.516...68G}
{Geach}, J.~E., {Hickox}, R.~C., {Diamond-Stanic}, A.~M., {et~al.} 2014, \nat,
  516, 68

\bibitem[{{Graham} \& {Langer}(1973)}]{1973ApJ...179..469G}
{Graham}, R., \& {Langer}, W.~D. 1973, \apj, 179, 469

\bibitem[{{Gray} \& {Scannapieco}(2010)}]{2010ApJ...718..417G}
{Gray}, W.~J., \& {Scannapieco}, E. 2010, \apj, 718, 417

\bibitem[{{Gregori} {et~al.}(1999){Gregori}, {Miniati}, {Ryu}, \&
  {Jones}}]{1999ApJ...527L.113G}
{Gregori}, G., {Miniati}, F., {Ryu}, D., \& {Jones}, T.~W. 1999, \apjl, 527,
  L113

\bibitem[{{Heckman}(2002)}]{2002ASPC..254..292H}
{Heckman}, T.~M. 2002, in Astronomical Society of the Pacific Conference
  Series, Vol. 254, Extragalactic Gas at Low Redshift, ed. J.~S. {Mulchaey} \&
  J.~T. {Stocke}, 292

\bibitem[{{Heckman} {et~al.}(1990){Heckman}, {Armus}, \&
  {Miley}}]{1990ApJS...74..833H}
{Heckman}, T.~M., {Armus}, L., \& {Miley}, G.~K. 1990, \apjs, 74, 833

\bibitem[{{Johansson} \& {Ziegler}(2013)}]{2013ApJ...766...45J}
{Johansson}, E.~P.~G., \& {Ziegler}, U. 2013, \apj, 766, 45

\bibitem[{{Kewley} \& {Ellison}(2008)}]{2008ApJ...681.1183K}
{Kewley}, L.~J., \& {Ellison}, S.~L. 2008, \apj, 681, 1183

\bibitem[{{Klein} {et~al.}(1994){Klein}, {McKee}, \&
  {Colella}}]{1994ApJ...420..213K}
{Klein}, R.~I., {McKee}, C.~F., \& {Colella}, P. 1994, \apj, 420, 213

\bibitem[{{Kornei} {et~al.}(2013){Kornei}, {Shapley}, {Martin}, {Coil}, {Lotz},
  \& {Weiner}}]{2013ApJ...774...50K}
{Kornei}, K.~A., {Shapley}, A.~E., {Martin}, C.~L., {et~al.} 2013, \apj, 774,
  50

\bibitem[{{Lacey}(1988)}]{1988ApJ...326..769L}
{Lacey}, C.~G. 1988, \apj, 326, 769

\bibitem[{{Lee}(2013)}]{2013JCoPh.243..269L}
{Lee}, D. 2013, Journal of Computational Physics, 243, 269

\bibitem[{{Li} {et~al.}(2013){Li}, {Frank}, \&
  {Blackman}}]{2013ApJ...774..133L}
{Li}, S., {Frank}, A., \& {Blackman}, E.~G. 2013, \apj, 774, 133

\bibitem[{{Mac Low} {et~al.}(1994){Mac Low}, {McKee}, {Klein}, {Stone}, \&
  {Norman}}]{1994ApJ...433..757M}
{Mac Low}, M.-M., {McKee}, C.~F., {Klein}, R.~I., {Stone}, J.~M., \& {Norman},
  M.~L. 1994, \apj, 433, 757

\bibitem[{{Mac Low} \& {Zahnle}(1994)}]{1994ApJ...434L..33M}
{Mac Low}, M.-M., \& {Zahnle}, K. 1994, \apjl, 434, L33

\bibitem[{{Mannucci} {et~al.}(2010){Mannucci}, {Cresci}, {Maiolino}, {Marconi},
  \& {Gnerucci}}]{2010MNRAS.408.2115M}
{Mannucci}, F., {Cresci}, G., {Maiolino}, R., {Marconi}, A., \& {Gnerucci}, A.
  2010, \mnras, 408, 2115

\bibitem[{{Marcolini} {et~al.}(2005){Marcolini}, {Strickland}, {D'Ercole},
  {Heckman}, \& {Hoopes}}]{2005MNRAS.362..626M}
{Marcolini}, A., {Strickland}, D.~K., {D'Ercole}, A., {Heckman}, T.~M., \&
  {Hoopes}, C.~G. 2005, \mnras, 362, 626

\bibitem[{{Martin}(1999)}]{1999ApJ...513..156M}
{Martin}, C.~L. 1999, \apj, 513, 156

\bibitem[{{Martin}(2006)}]{2006asup.book..337M}
---. 2006, {The Role of Galactic Winds in Galaxy Formation}, ed. J.~W. {Mason},
  337

\bibitem[{{Martin} {et~al.}(2015){Martin}, {Dijkstra}, {Henry}, {Soto},
  {Danforth}, \& {Wong}}]{2015ApJ...803....6M}
{Martin}, C.~L., {Dijkstra}, M., {Henry}, A., {et~al.} 2015, \apj, 803, 6

\bibitem[{{Martin} {et~al.}(2010){Martin}, {Scannapieco}, {Ellison}, {Hennawi},
  {Djorgovski}, \& {Fournier}}]{2010ApJ...721..174M}
{Martin}, C.~L., {Scannapieco}, E., {Ellison}, S.~L., {et~al.} 2010, \apj, 721,
  174

\bibitem[{{McCourt} {et~al.}(2015){McCourt}, {O'Leary}, {Madigan}, \&
  {Quataert}}]{2015MNRAS.449....2M}
{McCourt}, M., {O'Leary}, R.~M., {Madigan}, A.-M., \& {Quataert}, E. 2015,
  \mnras, 449, 2

\bibitem[{{McKee} \& {Begelman}(1990)}]{1990ApJ...358..392M}
{McKee}, C.~F., \& {Begelman}, M.~C. 1990, \apj, 358, 392

\bibitem[{{Mellema} {et~al.}(2002){Mellema}, {Kurk}, \&
  {R{\"o}ttgering}}]{2002A&A...395L..13M}
{Mellema}, G., {Kurk}, J.~D., \& {R{\"o}ttgering}, H.~J.~A. 2002, \aap, 395,
  L13

\bibitem[{{Morel}(2000)}]{2000JQSRT..65..769M}
{Morel}, J.~E. 2000, \jqsrt, 65, 769

\bibitem[{{Nipoti} \& {Binney}(2007)}]{2007MNRAS.382.1481N}
{Nipoti}, C., \& {Binney}, J. 2007, \mnras, 382, 1481

\bibitem[{{Oort} \& {Spitzer}(1955)}]{1955ApJ...121....6O}
{Oort}, J.~H., \& {Spitzer}, Jr., L. 1955, \apj, 121, 6

\bibitem[{{Orlando} {et~al.}(2008){Orlando}, {Bocchino}, {Reale}, {Peres}, \&
  {Pagano}}]{2008ApJ...678..274O}
{Orlando}, S., {Bocchino}, F., {Reale}, F., {Peres}, G., \& {Pagano}, P. 2008,
  \apj, 678, 274

\bibitem[{{Orlando} {et~al.}(2005){Orlando}, {Peres}, {Reale}, {Bocchino},
  {Rosner}, {Plewa}, \& {Siegel}}]{2005A&A...444..505O}
{Orlando}, S., {Peres}, G., {Reale}, F., {et~al.} 2005, \aap, 444, 505

\bibitem[{{Pan} {et~al.}(2012){Pan}, {Desch}, {Scannapieco}, \&
  {Timmes}}]{2012ApJ...756..102P}
{Pan}, L., {Desch}, S.~J., {Scannapieco}, E., \& {Timmes}, F.~X. 2012, \apj,
  756, 102

\bibitem[{{Peeples} {et~al.}(2014){Peeples}, {Werk}, {Tumlinson},
  {Oppenheimer}, {Prochaska}, {Katz}, \& {Weinberg}}]{2014ApJ...786...54P}
{Peeples}, M.~S., {Werk}, J.~K., {Tumlinson}, J., {et~al.} 2014, \apj, 786, 54

\bibitem[{{Pettini} {et~al.}(2001){Pettini}, {Shapley}, {Steidel}, {Cuby},
  {Dickinson}, {Moorwood}, {Adelberger}, \& {Giavalisco}}]{2001ApJ...554..981P}
{Pettini}, M., {Shapley}, A.~E., {Steidel}, C.~C., {et~al.} 2001, \apj, 554,
  981

\bibitem[{{Pichon} {et~al.}(2003){Pichon}, {Scannapieco}, {Aracil},
  {Petitjean}, {Aubert}, {Bergeron}, \& {Colombi}}]{2003ApJ...597L..97P}
{Pichon}, C., {Scannapieco}, E., {Aracil}, B., {et~al.} 2003, \apjl, 597, L97

\bibitem[{{Pittard} \& {Parkin}(2015)}]{2015arXiv151005478P}
{Pittard}, J.~M., \& {Parkin}, E.~R. 2015, ArXiv e-prints, arXiv:1510.05478

\bibitem[{{Rupke} {et~al.}(2005){Rupke}, {Veilleux}, \&
  {Sanders}}]{2005ApJS..160..115R}
{Rupke}, D.~S., {Veilleux}, S., \& {Sanders}, D.~B. 2005, \apjs, 160, 115

\bibitem[{{Scannapieco}(2013)}]{2013ApJ...763L..31S}
{Scannapieco}, E. 2013, \apjl, 763, L31

\bibitem[{{Scannapieco} \& {Br{\"u}ggen}(2015)}]{2015ApJ...805..158S}
{Scannapieco}, E., \& {Br{\"u}ggen}, M. 2015, \apj, 805, 158

\bibitem[{{Scannapieco} {et~al.}(2002){Scannapieco}, {Ferrara}, \&
  {Madau}}]{SFM02}
{Scannapieco}, E., {Ferrara}, A., \& {Madau}, P. 2002, \apj, 574, 590

\bibitem[{{Schwartz} \& {Martin}(2004)}]{2004ApJ...610..201S}
{Schwartz}, C.~M., \& {Martin}, C.~L. 2004, \apj, 610, 201

\bibitem[{{Shin} {et~al.}(2008){Shin}, {Stone}, \&
  {Snyder}}]{2008ApJ...680..336S}
{Shin}, M.-S., {Stone}, J.~M., \& {Snyder}, G.~F. 2008, \apj, 680, 336

\bibitem[{{Silich} {et~al.}(2003){Silich}, {Tenorio-Tagle}, \&
  {Mu{\~n}oz-Tu{\~n}{\'o}n}}]{2003ApJ...590..791S}
{Silich}, S., {Tenorio-Tagle}, G., \& {Mu{\~n}oz-Tu{\~n}{\'o}n}, C. 2003, \apj,
  590, 791

\bibitem[{{Thompson} {et~al.}(2016){Thompson}, {Quataert}, {Zhang}, \&
  {Weinberg}}]{2016MNRAS.455.1830T}
{Thompson}, T.~A., {Quataert}, E., {Zhang}, D., \& {Weinberg}, D.~H. 2016,
  \mnras, 455, 1830

\bibitem[{{Tremonti} {et~al.}(2004){Tremonti}, {Heckman}, {Kauffmann},
  {Brinchmann}, {Charlot}, {White}, {Seibert}, {Peng}, {Schlegel}, {Uomoto},
  {Fukugita}, \& {Brinkmann}}]{2004ApJ...613..898T}
{Tremonti}, C.~A., {Heckman}, T.~M., {Kauffmann}, G., {et~al.} 2004, \apj, 613,
  898

\bibitem[{{Turner} {et~al.}(2015){Turner}, {Schaye}, {Steidel}, {Rudie}, \&
  {Strom}}]{2015MNRAS.450.2067T}
{Turner}, M.~L., {Schaye}, J., {Steidel}, C.~C., {Rudie}, G.~C., \& {Strom},
  A.~L. 2015, \mnras, 450, 2067

\bibitem[{{Veilleux} {et~al.}(2005){Veilleux}, {Cecil}, \&
  {Bland-Hawthorn}}]{2005ARA&A..43..769V}
{Veilleux}, S., {Cecil}, G., \& {Bland-Hawthorn}, J. 2005, \araa, 43, 769

\bibitem[{{Veilleux} {et~al.}(2014){Veilleux}, {Teng}, {Rupke}, {Maiolino}, \&
  {Sturm}}]{2014ApJ...790..116V}
{Veilleux}, S., {Teng}, S.~H., {Rupke}, D.~S.~N., {Maiolino}, R., \& {Sturm},
  E. 2014, \apj, 790, 116

\bibitem[{{Vieser} \& {Hensler}(2007{\natexlab{a}})}]{2007A&A...475..251V}
{Vieser}, W., \& {Hensler}, G. 2007{\natexlab{a}}, \aap, 475, 251

\bibitem[{{Vieser} \& {Hensler}(2007{\natexlab{b}})}]{2007A&A...472..141V}
---. 2007{\natexlab{b}}, \aap, 472, 141

\bibitem[{{Wang}(1995)}]{1995ApJ...444..590W}
{Wang}, B. 1995, \apj, 444, 590

\bibitem[{{Wang} {et~al.}(2014){Wang}, {Nardini}, {Fabbiano}, {Karovska},
  {Elvis}, {Pellegrini}, {Max}, {Risaliti}, {U}, \&
  {Zezas}}]{2014ApJ...781...55W}
{Wang}, J., {Nardini}, E., {Fabbiano}, G., {et~al.} 2014, \apj, 781, 55

\bibitem[{{Weiner} {et~al.}(2009){Weiner}, {Coil}, {Prochaska}, {Newman},
  {Cooper}, {Bundy}, {Conselice}, {Dutton}, {Faber}, {Koo}, {Lotz}, {Rieke}, \&
  {Rubin}}]{2009ApJ...692..187W}
{Weiner}, B.~J., {Coil}, A.~L., {Prochaska}, J.~X., {et~al.} 2009, \apj, 692,
  187

\bibitem[{{Wiersma} {et~al.}(2009){Wiersma}, {Schaye}, \&
  {Smith}}]{2009MNRAS.393...99W}
{Wiersma}, R.~P.~C., {Schaye}, J., \& {Smith}, B.~D. 2009, \mnras, 393, 99

\bibitem[{{W{\"u}nsch} {et~al.}(2011){W{\"u}nsch}, {Silich}, {Palou{\v s}},
  {Tenorio-Tagle}, \& {Mu{\~n}oz-Tu{\~n}{\'o}n}}]{2011ApJ...740...75W}
{W{\"u}nsch}, R., {Silich}, S., {Palou{\v s}}, J., {Tenorio-Tagle}, G., \&
  {Mu{\~n}oz-Tu{\~n}{\'o}n}, C. 2011, \apj, 740, 75

\bibitem[{{Yirak} {et~al.}(2010){Yirak}, {Frank}, \&
  {Cunningham}}]{2010ApJ...722..412Y}
{Yirak}, K., {Frank}, A., \& {Cunningham}, A.~J. 2010, \apj, 722, 412

\bibitem[{{Zhang} {et~al.}(2015){Zhang}, {Thompson}, {Quataert}, \&
  {Murray}}]{2015arXiv150701951}
{Zhang}, D., {Thompson}, T.~A., {Quataert}, E., \& {Murray}, N. 2015, ArXiv
  e-prints, arXiv:1507.01951

\end{thebibliography}
\end{document}